\documentclass[twocolumn,showpacs,prb,amsmath,amssymb,nofootnoteinbib,superscriptaddress]{revtex4-2} 
\pdfoutput=1
\usepackage[dvipdfmx]{graphicx}
\usepackage{color}
\usepackage{hhline}
\usepackage{mathrsfs}
\usepackage{mathtools}
\usepackage{dcolumn}
\usepackage{bm}
\usepackage{multirow}
\usepackage{booktabs}
\usepackage{afterpage}
\usepackage{amsmath}
\usepackage{braket}
\usepackage{subfigure}

\allowdisplaybreaks

\begin{document}

\title{Improvement of the simplification method for the local two-particle full-vertex towards precise frequency behavior }

\author{Ryota Mizuno}\email{mizuno@presto.phys.sci.osaka-u.ac.jp}
\affiliation{Forefront Research Center, Osaka University, 1-1 Machikaneyama, Toyonaka, Osaka 560-0043, Japan}
\author{Kazuhiko Kuroki}
\affiliation{Department of Physics, Osaka University, 1-1 Machikaneyama, Toyonaka, Osaka 560-0043, Japan}
\author{Masayuki Ochi}
\affiliation{Forefront Research Center, Osaka University, 1-1 Machikaneyama, Toyonaka, Osaka 560-0043, Japan}
\affiliation{Department of Physics, Osaka University, 1-1 Machikaneyama, Toyonaka, Osaka 560-0043, Japan}

\date{\today}
\begin{abstract}
Estimating the local two-particle vertex functions, which are crucial for capturing the spatial fluctuation of the effective field beyond the single-site DMFT, is still challenging.
In our previous work~\cite{doi:10.7566/JPSJ.91.034002}, we developed a computationally efficient method for estimating the local full-vertex in DMFT, where we can obtain the local two-particle full-vertex from the one-particle self-energy. 
In this study, we further enhance our method by refining its formulation to be more faithful to the diagrammatic structure of the full-vertex.
With this improvement, we can qualitatively reproduce the characteristic frequency structures of the full-vertex obtained by the numerically exact methods.
In particular, the improved version of the simplified full-vertex captures a sharp value change in the cross structure.

\end{abstract}

\maketitle

\section{Introduction}

The dynamical mean field theory (DMFT)~\cite{RevModPhys.68.13} is one of the powerful methods for analyzing strongly correlated systems.
DMFT can correctly capture the strong local correlation effects and, therefore, has advantages in describing some aspects of the Kondo physics and the metal-to-Mott insulator transition.
On the other hand, the non-local correlations ignored in DMFT are essential for low dimensional systems and at low temperatures, where various fascinating phenomena, such as anisotropic superconductivity and the pseudo-gap, emerge.

Several extensions of DMFT for considering the non-local correlations have been suggested so far. 
There are roughly two types of extensions: the cluster type~\cite{RevModPhys.77.1027} and the diagrammatic type~\cite{RevModPhys.90.025003}. 
In the cluster-type extensions, we do not have to calculate the local two-particle quantities. 
In contrast, the system size we can take is strictly limited due to the rapid increase of the numerical cost, so the non-local correlations we can consider are limited to the short-range one.
In the diagrammatic extensions, on the other hand, we can take a sufficiently large system size and the long-range non-local correlations. 
For example, we can analyze the superconducting fluctuations or pseudo-gap phenomena in the dual fermion method~\cite{PhysRevB.77.033101,PhysRevB.79.045133, PhysRevB.90.235132,PhysRevB.97.115150,PhysRevB.98.155117} and dynamical vertex approximation~(D$\Gamma$A)~\cite{PhysRevB.75.045118,andp.201100036,doi:10.1143/JPSJ.75.054713}, where the local two-particle vertex functions obtained by single-site DMFT are important building blocks of calculations in obtaining the non-local fluctuation.  
However, estimating the local two-particle quantities with numerically exact impurity solvers such as the continuous-time quantum Monte Carlo (CT-QMC)~\cite{PhysRevB.72.035122,PhysRevLett.97.076405,PhysRevB.76.235123,PhysRevB.74.155107,doi:10.1143/JPSJ.76.114707}, the exact diagonalization (ED)~\cite{PhysRevLett.72.1545,PhysRevB.86.165128}, or the numerical renormalization group~\cite{PhysRevX.11.041006} is still challenging, especially in multi-band systems, although several efforts have been carried out~\cite{PhysRevB.94.125153,10.21468/SciPostPhys.8.1.012,Moghadas2024,PhysRevB.109.115128}.

In our previous study~\cite{doi:10.7566/JPSJ.91.034002}, we developed a method for simplifying the local full-vertex to overcome these practical difficulties in the diagrammatic extensions. 
We showed that we can approximate the local full-vertex in a simple form based on its frequency and diagrammatic structures.
We can obtain the simplified full-vertex that can capture the essential frequency structures of the numerically exact full-vertex with a very low numerical cost.
However, the simplified full-vertex can have qualitatively inconsistent frequency behavior with the exact one, depending on the situation.

In this study, we improve the simplification method for the local two-particle full-vertex by refining its formulation to be more faithful to the diagrammatic structure. 
We also improve the method for obtaining the simplified two-particle full-vertex from the one-particle self-energy, called the self-energy to full-vertex (S2F) method, to apply the improved version of the simplified full-vertex.
We can obtain the full-vertex qualitatively consistent with that obtained by the exact impurity solvers~\cite{PhysRevB.96.035114,PhysRevB.86.125114} with significantly low numerical cost.

This paper is organized as follows. 
In Sect.~\ref{sec:2024-07-11-13-12}, we introduce the model and the two-particle Green's function. 
In Sect.~\ref{sec:2024-07-11-13-13}, we introduce the basic concept of the simplification of the local two-particle full-vertex and describe the improvement of the simplification of the full-vertex and the S2F method. 
We show the results in Sect.~\ref{sec:2024-07-07-15-52}. 
The discussion is presented in Sect.~\ref{sec:2024-07-11-13-14}.
The conclusion is given in Sect.~\ref{sec:2024-07-11-13-15}.

\section{Model and Two-Particle Green's Function}\label{sec:2024-07-11-13-12}

\begin{figure*}[t]
  \centering
  {\includegraphics[width=180mm,clip]{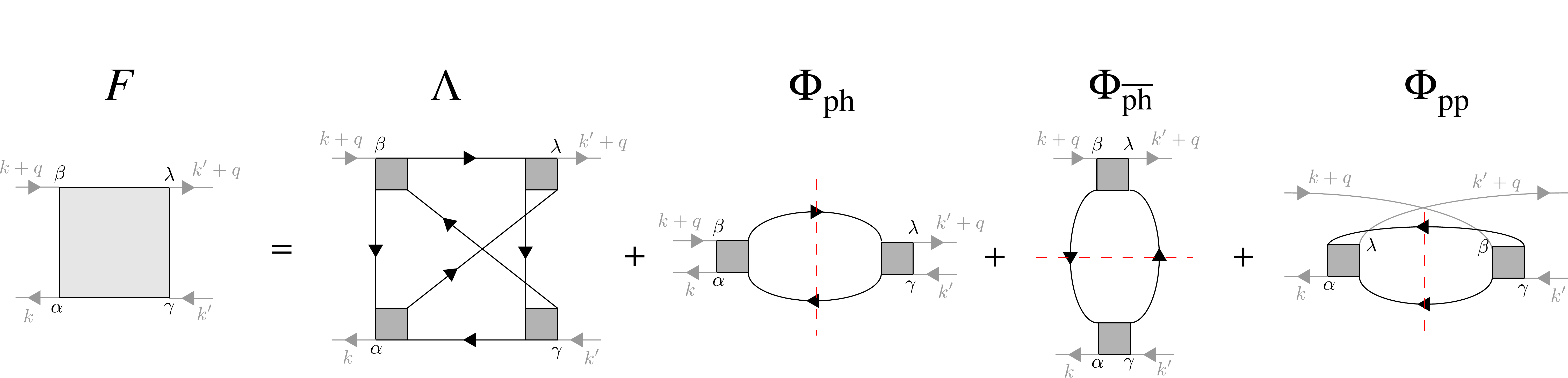}} 
  \caption{The decomposition of the full vertex.
    The full vertex can be divided into four parts:
    the fully irreducible part ($\Lambda$) and the reducible parts ($\Phi_{l}$,  $l=$ ph, ${\rm \overline{ph}}$, pp).  
    The reducible vertices $\Phi_{l}$ can be divided into two parts by cutting a pair of Green's functions as indicated by the red dashed lines.
  }
  \label{fig:2020-05-09-22-21}
\end{figure*}

We consider the multiband Hubbard model described by the following Hamiltonian
\begin{align}
  H 
  =&
  \sum_{ij}\sum_{\alpha\beta}t_{ij,\alpha\beta}c^{\dagger}_{i\alpha}c_{j\beta} 
  +
  \dfrac{1}{4} \sum_{i} \sum_{\alpha\beta\gamma\lambda} U_{\alpha\beta\gamma\lambda} c^{\dagger}_{i\alpha}c^{\dagger}_{i\lambda}c_{i\gamma}c_{i\beta}, 
  \label{eq:2024-05-09-14-07}
\end{align}
where 
the subscripts with Roman letters indicate unit cells 
and
Greek letters the set of the degrees of freedom of spin, orbital, and site.
$t_{ij,\alpha\beta}$ is the hopping integral 
and 
$U_{\alpha\beta\gamma\lambda}$ is the Coulomb repulsion.
$c_{i\alpha}^{(\dagger)}$ is the annihilation (creation) operator.

In the presence of the translational invariance in time and space, we can write down the two-particle Green's function in the momentum space as follows
\begin{align}
  G^{(2)}_{\alpha\beta\gamma\lambda}&(\bm{k},\bm{k}',\bm{q}, \tau_{1},\tau_{2},\tau_{3}) 
  \nonumber \\
  &=
  \Bigl< T c_{\bm{k}\alpha}(\tau_{1})c^{\dagger}_{\bm{k}+\bm{q}\beta}(\tau_{2})c_{\bm{k}'+\bm{q}\lambda}(\tau_{3})c^{\dagger}_{\bm{k}'\gamma} \Bigr>,
  \label{eq:2020-10-08-23-02}
\end{align}
where
$c^{(\dagger)}(\tau)=e^{\tau H}c^{(\dagger)}e^{-\tau H}$ is the Heisenberg representation of annihilation (creation) operator.
Fourier transformation is given by 
\begin{align}
  G^{(2)}&(\bm{k},\bm{k}',\bm{q}, \tau_{1},\tau_{2},\tau_{3}) \nonumber \\
  =&
  \dfrac{1}{\beta^{3}} \sum_{nn'm} G^{(2)}(\bm{k},\bm{k}',\bm{q}, i\omega_{n},i\omega_{n'},i\nu_{m})
  \nonumber \\
  &\times e^{-i\omega_{n} \tau_{1}} e^{i(\omega_{n}+\nu_{m})\tau_{2}} e^{-i(\omega_{n'}+\nu_{m})\tau_{3}},
  \label{eq:2020-10-08-23-03}
\end{align}
where 
$\omega_{n}=(2n+1)\pi T$ and  $\nu_{m}=2m \pi T$ with $n, m \in {\mathbb Z}$ are the fermionic and bosonic Matsubara frequencies, respectively.
Hereafter, we may omit the integer variables as $\omega_{n} \to \omega, \ \omega_{n'} \to \omega', \ \nu_{m} \to \nu$ without notice, if there is no confusion.
Generally, we can write down the two-particle Green's function in the following form
\begin{align}
  &G^{(2)}_{\alpha\beta\gamma\lambda}(k,k',q) \nonumber \\
  &=
  G_{\alpha\beta}(k) G_{\lambda\gamma}(k') \delta_{q,0} 
  -
  G_{\alpha\gamma}(k) G_{\lambda\beta}(k+q) \delta_{kk'}
  \nonumber \\
  &-
  \hspace{-5pt}\sum_{\alpha'\beta'\gamma'\lambda'}\hspace{-5pt}
  G_{\alpha\gamma'}(k) G_{\lambda'\beta}(k+q) F_{\gamma'\lambda'\alpha'\beta'}(k,k',q) \nonumber \\
  &\times G_{\alpha'\gamma}(k') G_{\lambda\beta'}(k'+q),
  \label{eq:2024-05-09-14-22}
\end{align}
where  
$k=(\bm{k},i\omega_{n})$ and $q=(\bm{q},i\nu_{m})$ denote the generalized fermionic and bosonic momenta, respectively.
$F$ is called the full-vertex.
The term containing $F$ is called the connected term, and the others are called disconnected.
When we consider the diagrammatic structure of the full-vertex $F$, we define the irreducible susceptibilities concerning the following three channels (${\rm ph, \overline{ph}, pp }$):
\begin{align}
  \chi_{0,\alpha\beta\gamma\lambda}(k,k',q) =& 
  \begin{cases}
    - G_{\alpha\gamma}(k)G_{\lambda\beta}(k+q) \delta_{kk'} \hspace{5pt} &(\text{ph}) \\
    G_{\alpha\beta}(k) G_{\lambda\gamma}(k') \delta_{q0}    &({\overline{\rm ph}}) \\
    G_{\alpha\gamma}(k)G_{\beta\lambda}(-k-q)\delta_{kk'}     &(\text{pp})
  \end{cases}.
  \label{eq:2024-05-09-14-46} 
\end{align}
The full vertex $F$ can be divided into four parts
\begin{align}
  F(D) &= \Lambda(D) + \Phi_{\rm ph}(D) + \Phi_{\rm \overline{ph}}(D) + \Phi_{\rm pp}(P) \nonumber \\
  &= \Lambda(D) + \Phi_{\rm ph}(D) - \Phi_{\rm ph}(C) + \Phi_{\rm pp}(P),
  \label{eq:2020-05-09-22-20}
\end{align}
where 
$\Phi_{l}$ $(l={\rm ph,\overline{ph},pp})$
is the set of reducible diagrams in channel $l$,
and 
$\Lambda$ is the set of fully irreducible diagrams. 
Also, we introduce the notation for the set of orbital and frequency variables here as
\begin{align}
  D =& (\alpha,\beta,\gamma,\lambda), (k,k',q), \label{eq:2020-05-10-14-49} \\
  T =& (\alpha,\beta,\lambda,\gamma), (k,-q-k',q), \label{eq:2021-08-12-14-54} \\ 
  C =& (\alpha,\gamma,\beta,\lambda), (k,k+q,k'-k), \label{eq:2020-05-10-14-51} \\
  P =& (\alpha,\lambda,\gamma,\beta), (k,k',-q-k-k'), \label{eq:2020-05-10-14-52} \\
  X =& (\alpha,\gamma,\lambda,\beta), (k,-k-q,k'-k). \label{eq:2020-05-10-14-53}
\end{align}
The diagrammatic representation is shown in Fig.~\ref{fig:2020-05-09-22-21}.
Since no diagram satisfies reducibility in two or more channels simultaneously, we can write
\begin{align}
  F =& \Gamma_{l} + \Phi_{l}, 
  \label{eq:2020-05-09-22-53} \\
  \Gamma_{l} =& \Lambda + \Phi_{l_{1}} + \Phi_{l_{2}} \hspace{10pt} (l\neq l_{1} \neq l_{2}), 
  \label{eq:2020-05-09-22-54} \\
  \Phi_{l} =& 
  -\Gamma_{l}\chi_{0} F = -\Gamma_{l}\chi_{l} \Gamma_{l}, 
\end{align}
where 
$\Gamma_{l}$ is the set of diagrams irreducible in channel $l$ 
and 
is called the irreducible vertex in $l$.
The generalized susceptibility in channel $l$ is given by
\begin{align}
  \chi_{l} =& \chi_{0} - \chi_{0}\Gamma_{l}\chi_{l} = \chi_{0} - \chi_{0} F\chi_{0} . \label{eq:2020-05-12-14-40} 
\end{align}
We omit the orbital and frequency index in Eqs.~(\ref{eq:2020-05-09-22-53})-(\ref{eq:2020-05-12-14-40}) since it depends on the channels.
Hereafter, we may omit the index when we do not specify the channel.
From 
Eqs.~(\ref{eq:2020-05-09-22-53}) to (\ref{eq:2020-05-12-14-40}), which are called the parquet equations~\cite{e_023_03_0489,PhysRevB.86.125114,Janis_1998,PhysRevB.60.11345}, 
we can calculate $F$ exactly if we know the exact $\Lambda$.
However, it is almost impossible to obtain the exact $\Lambda$ 
and the procedure to obtain $\Phi_{l}$ is computationally very expensive. 
Thus, 
some approximations or simplifications have been proposed~\cite{doi:10.1143/JPSJ.79.094707,PhysRevB.75.165108,PhysRevB.83.035114, PhysRevB.104.035160}.

\section{Simplification of the Local Full-vertex} \label{sec:2024-07-11-13-13}
In our previous study~\cite{doi:10.7566/JPSJ.91.034002}, we proposed a simplification method for the local full-vertex based on the diagrammatic structures of two-particle vertices.
In this work, we improve the simplified form of the full-vertex by refining its formulation to be more faithful to the diagrammatic structure.
In this section, we briefly explain the basic concept of the simplification and then show how we improved the formulation from the previous one.

\

\subsection{Basic Concept of the Simplification of the Local Full-vertex} \label{sec:2024-07-01-14-44}

\begin{figure}[t]
  \centering
  {\includegraphics[width=80mm,clip]{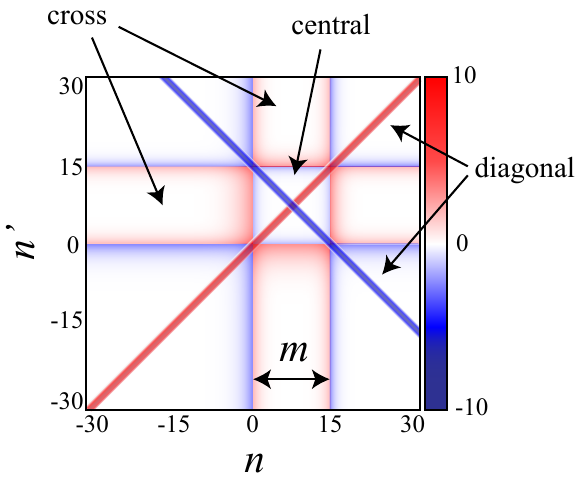}} 
  \caption{  
    A schematic figure of $(\omega_{n},\omega_{n'})$ plane of the local full-vertex at $\nu=30\pi T \ (m=15)$.
    (i) The $\omega-\omega'=0$ and $\omega+\omega'+\nu=0$ lines, (ii) the band sandwiched between $\omega^{(\prime)}=0$ and $\omega^{(\prime)}+\nu=0$ lines, and the square with side length $\nu$ centered at the $\omega+\nu/2 = \omega^{\prime}+\nu/2=0$ point are called (i) diagonal, (ii) cross, and (iii) central structures, respectively.
}
  \label{fig:2024-05-09-17-13}
\end{figure}

In the presence of the translational invariance,
the local two-particle full-vertex depends on three frequencies: two fermionic $\omega,\omega'$, and one bosonic $\nu$.
In the $(\omega,\omega')$ plane, the full-vertex has three characteristic structures, like a schematic figure of the local full-vertex in Fig.~\ref{fig:2024-05-09-17-13}.
The full-vertex largely changes its values near (i) $\omega-\omega'=0$ and $\omega+\omega'+\nu=0$ lines, (ii) the band sandwiched between $\omega^{(\prime)}=0$ and $\omega^{(\prime)}+\nu=0$ lines, and
(iii) the square with side length $\nu$ centered at $\omega+\nu/2 = \omega'+\nu/2=0$ point.
We call these structures (i) diagonal, (ii) cross, and (iii) central, respectively.

\begin{figure*}[t]
  \centering
  {\includegraphics[width=160mm,clip]{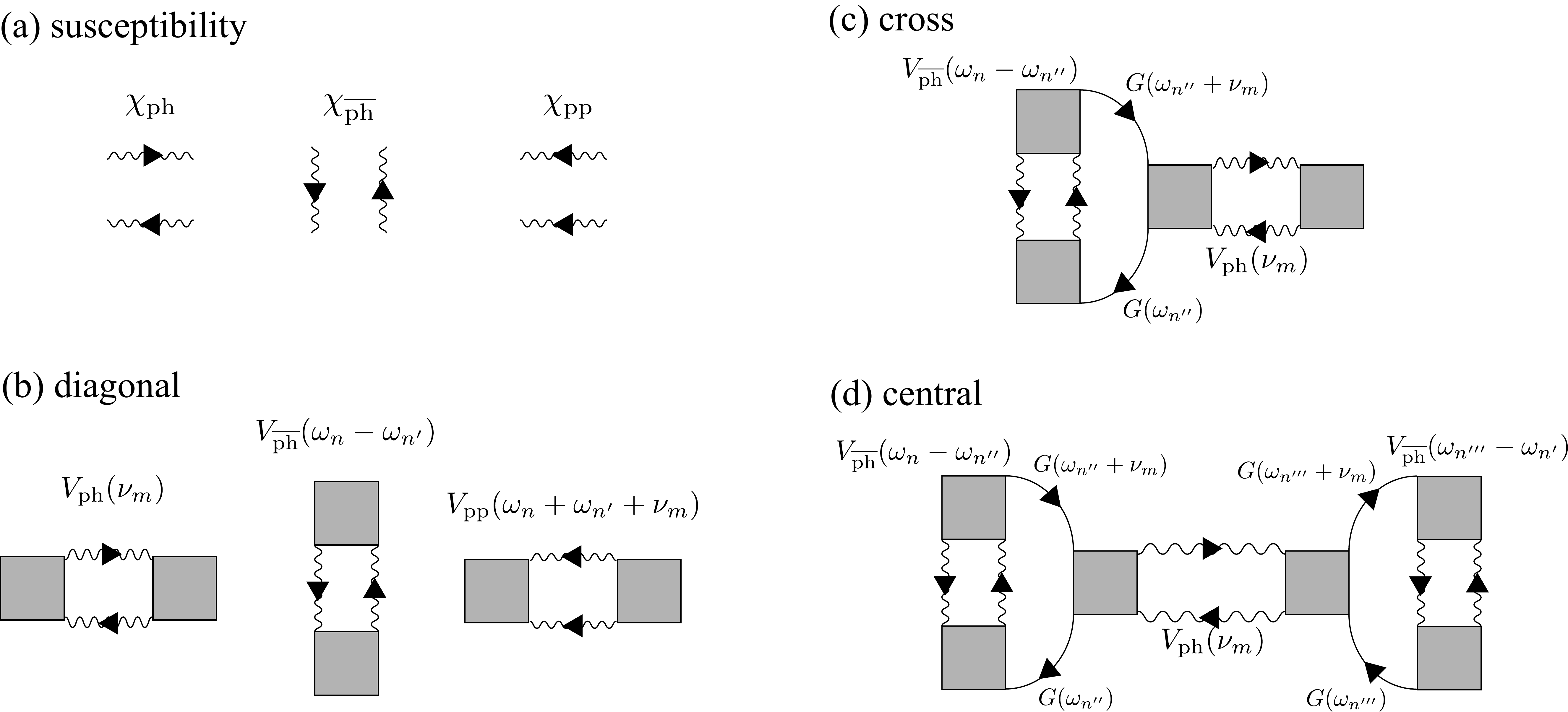}} 
  \caption{  
    Diagrams of (a)~susceptibilities and (b-d)~vertices mainly contribute to (b)~diagonal, (c)~cross and (d)~central structures.
  }
  \label{fig:2024-05-09-23-10}
\end{figure*}

We show diagrams that mainly contribute to these three characteristic frequency structures in Fig.~\ref{fig:2024-05-09-23-10}.
Here, we express the generalized susceptibilities  in ph, $\overline{\text{ph}}$, and pp channels with a pair of wavy lines as (a), and we consider here the vertices represented by squares in (b)-(d) to be constant vertices for simplicity
\footnote{
   Although the vertices represented by squares depend on the three frequencies, we can expect that their constant part mainly contributes to the global frequency structure of the full-vertex we focus on in this study.
In other words, since they have frequency dependencies only near the origin of $(\omega, \omega', \nu)$-space, their frequency-dependent parts have contributions only in the vicinity of the origin, and have little effect on the global frequency structure of the full-vertex.
}
.
Although the susceptibilities in (a) depend on three frequencies, vertices in (b) have only the bosonic frequency since they lost the other two frequency dependencies when we integrated the internal frequency variables as follows
\begin{align}
  &\sum_{n'',n'''}V_{1}(\omega,\omega'',\nu)\chi(\omega'',\omega''',\nu)V_{2}(\omega''',\omega',\nu) \nonumber \\
  &\xrightarrow{\text{$V_{1}$, $V_{2}$ = const.}} 
  V_{1} \Bigl[\sum_{n'',n'''}\chi(\omega'',\omega''',\nu)\Bigr] V_{2} = V^{\text{(b)}}_{\text{ph}}(\nu).
  \label{eq:2024-07-25-16-28}
\end{align}
We can obtain the frequency dependence of the (b)-type vertices in ${\overline{\text{ph}}}$ and pp channels as  $V^{\text{(b)}}_{\overline{\text{ph}}}(\omega-\omega')$ and $V^{\text{(b)}}_{\text{pp}}(-\nu-\omega-\omega')$ by transforming $D\to C$ and $D \to P$ in Eq.~(\ref{eq:2020-05-10-14-49})-(\ref{eq:2020-05-10-14-53}).
These vertices that convey only the bosonic frequency, which is closely related to the two-particle fluctuation, such as spin, charge, and orbital fluctuation, give the diagonal structure. 
For example, the diagonal structures in the particle-particle channel mentioned above are the main contributors to the superconducting phase transition, and the diagonal structures of the spin channel mentioned later are those to the magnetic phase transitions.

The cross and central structures come from diagrams shown in Figs.~\ref{fig:2024-05-09-23-10}~(c) and \ref{fig:2024-05-09-23-10}~(d), respectively. 
When two (b)-type vertices from different channels combine as Fig.~\ref{fig:2024-05-09-23-10}~(c), the resulting vertex loses the $\omega$ or $\omega'$ dependence due to the internal variable integration, for example in ph channel, 
\begin{align}
  V^{\text{(c)}}_{\text{ph}}&(\omega,\nu) \nonumber \\
  =&
  T \sum_{n''} V^{\text{(b)}}_{\overline{\text{ph}}}(\omega-\omega'')G(\omega''+\nu)G(\omega'')V^{\text{(b)}}_{\text{ph}}(\nu) \nonumber \\
  =&
  u_{1}(\omega,\nu)V^{\text{(b)}}_{\text{ph}}(\nu),
  \label{eq:2024-08-16-14-19}
\end{align}
where $u_{1}=T\sum V_{\overline{\text{ph}}}^{\text{(b)}}GG$.
This type of vertices gives the cross structure.
Similarly, when three (b)-type vertices combine as in (d), the resulting vertex depends on $\omega$ and $\omega'$ independently as
\begin{align}
  V^{\text{(d)}}_{\text{ph}}&(\omega,\omega',\nu) \nonumber \\
  =&
  T^{2}\sum_{n'',n'''}
  V^{\text{(b)}}_{\overline{\text{ph}}}(\omega-\omega'')G(\omega''+\nu)G(\omega'')
  \nonumber \\
  &\times V^{\text{(b)}}_{\text{ph}}(\nu)
  G(\omega'''+\nu)G(\omega''')
  V^{\text{(b)}}_{\overline{\text{ph}}}(\omega'''-\omega') \nonumber \\
  =&
  u_{1}(\omega,\nu)V_{\text{ph}}^{\text{(b)}}(\nu) u_{3}(\omega',\nu),
  \label{eq:2024-08-16-15-20}
\end{align}
where $u_{1}=T\sum V_{\overline{\text{ph}}}^{\text{(b)}}GG$ and $u_{2}=T\sum GG V_{\overline{\text{ph}}}^{\text{(b)}}$.
This type of vertices gives the central structure.
Growth of the cross and central structures leads to developing the low-energy part of the one-particle self-energy. 
Therefore, these structures are crucial in a strongly correlated regime and closely related to the Mott transition.
One can also understand this from their diagrammatic origins; namely, the cross and central structures come from higher-order diagrams than the diagonal structure.

The complex frequency dependence of the full-vertex prevents us from evaluating it and brings disadvantages in terms of the numerical costs in the diagrammatic extensions of DMFT, which requires the local two-particle vertex functions.
In our previous study~\cite{doi:10.7566/JPSJ.91.034002}, we proposed a simplification of the full-vertex to circumvent these difficulties while keeping the indispensable frequency structures (diagonal, cross, and central).
As shown in Eqs.~(\ref{eq:2024-08-16-14-19}) and (\ref{eq:2024-08-16-15-20}), the (c) and (d) type vertices, which mainly contribute to the cross and central structures, consist of (b)-type vertices $V^{\text{(b)}}$ and the parts $u_{1}$ and $u_{2}$. 
Furthermore, $u_{1}$ and $u_{2}$ can be approximated to the simpler form below. 
$V^{\text{(b)}}_{l}(\nu) \ (l=\text{ph},\overline{\text{ph}}, \text{pp})$ has a large value in the vicinity of $\nu=0$.
Especially when the fluctuation that $V^{\text{(b)}}_{l}$ conveys becomes essential, $V^{\text{(b)}}_{l}(\nu=0)$ becomes extremely large.
Hence, $V^{\text{(b)}}_{l}$ acts like the delta function or Gaussian in the frequency integration in Eqs.~(\ref{eq:2024-08-16-14-19}) and (\ref{eq:2024-08-16-15-20}).
Therefore, we can expect that the frequency dependence of the cross and central structures reflects that of the one-particle Green's functions which connect two $V^{\text{(b)}}_{l}$'s, namely the Green's functions in $u_{1}$ and $u_{2}$ in Eqs.~(\ref{eq:2024-08-16-14-19}) and (\ref{eq:2024-08-16-15-20}).
Then, we can approximately separate the variables of $u_{1}$ and $u_{2}$ as
\begin{align}
  u_{1}(\omega,\nu) \approx& C(\omega)C(\omega+\nu), 
  \label{eq:2024-08-16-17-26} \\
  u_{2}(\omega',\nu) \approx& C(\omega')C(\omega'+\nu).
  \label{eq:2024-08-16-17-27} 
\end{align}
We can apply a similar approximation to the vertices in $\overline{\text{ph}}$ and pp channels. 
Namely, similarly to the ph channel in Eqs.~(\ref{eq:2024-08-16-14-19})-(\ref{eq:2024-08-16-17-27}), the vertices that mainly give the cross and central structures in the $\overline{\text{ph}}$ and pp channels can be approximated to the product form of the (b)-type vertices and the factor $C$ that depends on a fermionic frequency.
Given this approximation, 
in our previous study, 
we proposed the following simplification form of the full-vertex.
\begin{align} 
  F(\omega&,\omega',\nu) \nonumber \\
  \approx&
  \tilde{C}(\omega)\tilde{C}(\omega+\nu)F_{0}(\omega,\omega',\nu)
  \tilde{C}(\omega')\tilde{C}(\omega'+\nu), 
  \label{eq:2021-04-14-21-16} \\
  F_{0}(\omega&,\omega',\nu) \nonumber \\
  =&
  \Lambda + \Phi_{\rm ph}(\nu) 
  + \Phi_{\rm \overline{ph}}(\omega-\omega') + \Phi_{\rm pp}(\omega+\omega'+\nu),
  \label{eq:2021-04-14-21-17}
\end{align}
where $\tilde{C}$ gives the frequency dependence of the cross and central structures and $F_{0}$ constant and diagonal structures.
We should note that $\tilde{C}$ (not $C$) here corresponds to $C$ in our previous paper~\cite{doi:10.7566/JPSJ.91.034002}.
The full-vertex in Eqs.~(\ref{eq:2021-04-14-21-16}) and (\ref{eq:2021-04-14-21-17}) include all the constant vertices, (b)-type vertices and the simplified form of (c),(d)-type vertices.
Therefore, this full-vertex can roughly reproduce the three characteristic structures of the exact full-vertex.

\subsection{Improved version of the simplification form of the full-vertex} \label{sec:2024-10-23-14-21}

\begin{figure} 
  \centering
  {\includegraphics[width=85mm,clip]{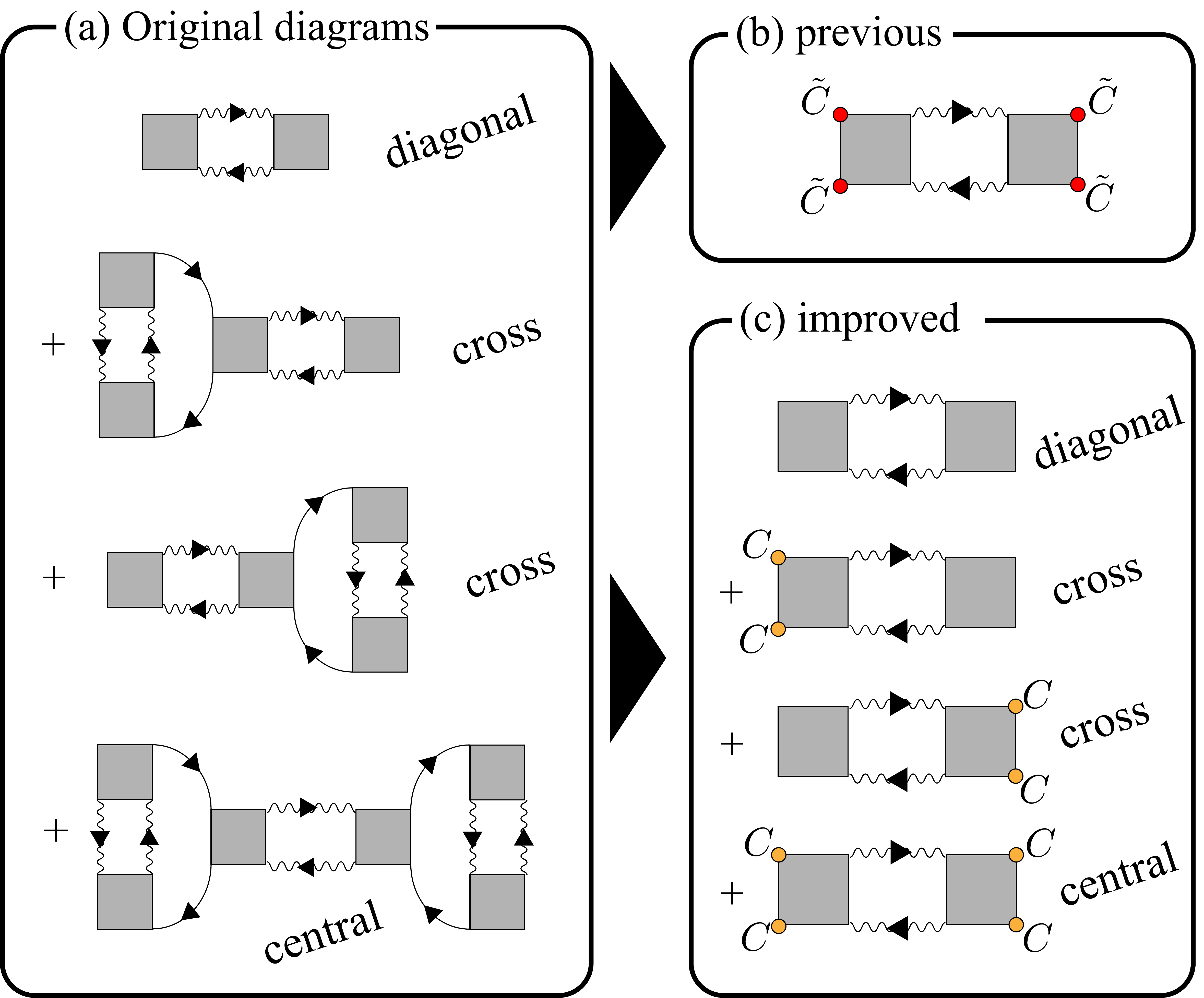}} 
  \caption{  
    A schematic of the improvement of the simplification of the vertices. 
    (a) Examples of diagrams that mainly construct the three characteristic structures.
    (b) The diagrammatic expression of the previous version of the simplified vertex. 
    (c) The diagrammatic expression of the improved version of the simplified vertex.
    The diagrams in (a) are expressed in one expression taking all the diagrams together in the previous version (b), while they are expressed separately depending on the frequency structures in the improved version (c).  
}
  \label{fig:2024-10-15-16-00}
\end{figure}

As we show later in Sect.~\ref{sec:2024-07-07-15-52}, the simplified full-vertex in Eqs.~(\ref{eq:2021-04-14-21-16}) and (\ref{eq:2021-04-14-21-17}) can roughly reproduce the three characteristic frequency structures of the exact full-vertex. 
However, the cross and central structures of the simplified full-vertex show qualitatively inconsistent behavior with that of the exact full-vertex.
This deviation mainly comes from the drawback that the simplified full-vertex in Eqs.~(\ref{eq:2021-04-14-21-16}) and (\ref{eq:2021-04-14-21-17}) has extra terms that can not appear in a diagrammatic manner.
In this section, we propose an improved version of the full-vertex simplification that is more faithful to the diagrammatic structure than the previous one.
Fig.~\ref{fig:2024-10-15-16-00} shows a schematic of our improvement of the simplification.
In the previous version, we express the sum of the diagrams that give the three frequency structures with one expression as in Fig.~\ref{fig:2024-10-15-16-00}~(b).
We refine the formulation so that they can have separate simplified expressions depending on the frequency structures as in Fig.~\ref{fig:2024-10-15-16-00}~(c). 
We explain the details below.

The relations among two-particle vertices are as follows
\begin{align}
  F &= \Lambda + \Phi_{l} + \Phi_{l_{1}} + \Phi_{l_{2}} , \hspace{20pt} (l\neq l_{1} \neq l_{2})
  \nonumber \\
  &= \Lambda + \Phi_{l} + \gamma_{l} 
  \nonumber \\
  &= \Gamma_{l} + \Phi_{l},
\end{align}
where $\Lambda$, $\Phi_{l}$, and $\Gamma_{l}$ are the fully-irreducible-vertex, reducible-vertex, and irreducible-vertex in $l$ channel, respectively.
Here, we define $\gamma_{l}\equiv \Phi_{l_{1}} + \Phi_{l_{2}}$ and so  $\Gamma_{l} = \Lambda + \gamma_{l}$. 
By using these equations, we can obtain
\begin{align}
  \Phi_{l}
  =& -\Gamma_{l}\chi_{l}\Gamma_{l} 
  \nonumber \\
  =& -(\Lambda + \gamma_{l}) \chi_{l} (\Lambda + \gamma_{l})
  \nonumber \\
  =&
  -\Lambda \chi_{l} \Lambda 
  - \gamma_{l}\chi_{l}\Lambda
  - \Lambda\chi_{l}\gamma_{l}
  - \gamma_{l}\chi_{l}\gamma_{l}.
  \label{eq:2024-03-22-00-52}
\end{align}
The first term in the last row in Eq.~(\ref{eq:2024-03-22-00-52}) contains the (b)-type diagrams in Fig~\ref{fig:2024-05-09-23-10}, 
the second and third terms (c)-type, and the last term (d)-type. 
To extract the structure where a pair of Green's functions connects the two channels of vertices, we further transform the reducible vertex as follows using the relation $\chi_{l} = \chi_{0} - \chi_{0}\Gamma_{l}\chi_{l}$, 
\begin{align}
  \Phi_{l}
  =&
  (-\gamma_{l}\chi_{0}\Gamma_{l}\Lambda^{-1}) \Lambda 
  +  \Lambda  (-\Lambda^{-1}\Gamma_{l}\chi_{0}\gamma_{l}) \nonumber \\
  & + (-\gamma_{l}\chi_{0}\Gamma_{l}\Lambda^{-1}) \Lambda  (-\Lambda^{-1}\Gamma_{l}\chi_{0}\gamma_{l}) \nonumber \\
  & + \Psi_{l}
  + (-\gamma_{l}\chi_{0}\Gamma_{l}\Lambda^{-1}) \Psi_{l}
  +  \Psi_{l}  (-\Lambda^{-1}\Gamma_{l}\chi_{0}\gamma_{l}) \nonumber \\
  &+ (-\gamma_{l}\chi_{0}\Gamma_{l}\Lambda^{-1}) \Psi_{l}  (-\Lambda^{-1}\Gamma_{l}\chi_{0}\gamma_{l})
  \nonumber \\
  & + 
  (\Gamma_{l}-\Lambda) [ \chi_{0} -  \chi_{0}(\Gamma_{l}\Lambda^{-1}-1) \Gamma_{l}\chi_{0} ] (\Gamma_{l}-\Lambda),
  \label{eq:2024-05-19-14-46}
\end{align}
where, $\Psi_{l} = -\Lambda\chi_{l}\Lambda$ [see Appendix.~\ref{sec:2024-11-12-16-03} for details].

\begin{figure*}[t]
  \centering
  {\includegraphics[width=140mm,clip]{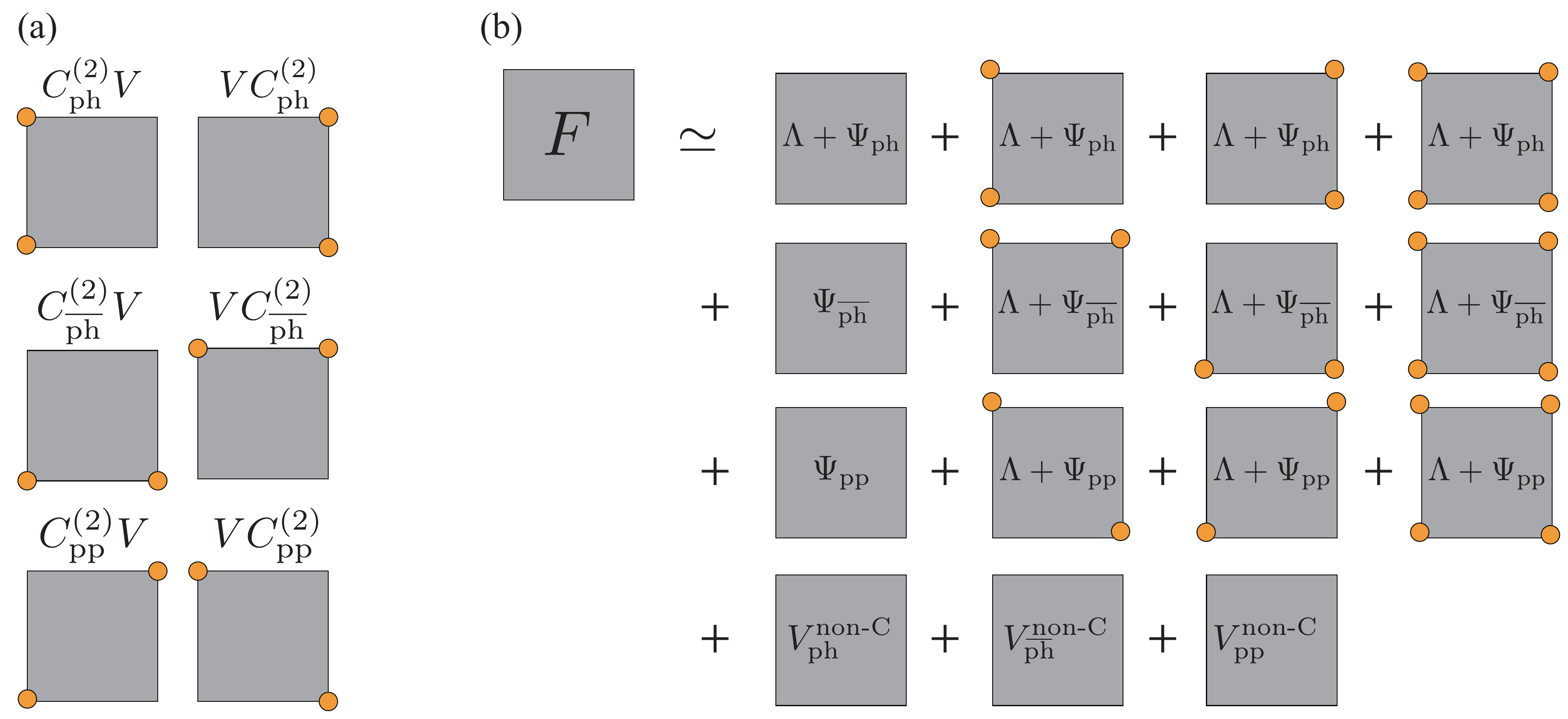}} 
  \caption{  
    The diagrammatic expression of (a)~how the correction factors $C$'s are attached to the vertices and (b)~the simplified full-vertex. 
  }
  \label{fig:2024-05-26-15-48}
\end{figure*}

Now, we introduce some approximations. 
We replace the fully-irreducible vertex $\Lambda$ and the irreducible vertex $\Gamma_{l}$ with the bare vertex and the frequency-independent one obtained by simplified parquet~\cite{doi:10.1143/JPSJ.79.094707,PhysRevB.104.035160,doi:10.7566/JPSJ.91.034002}, respectively. 
Then, the generalized susceptibility $\chi_{l}$ becomes independent of the two fermionic frequencies. 
Since $\gamma_{l}$ is the (b)-type vertex in channels other than $l$ and $\chi_{0} = -GG$, $-\gamma_{l}\chi_{0}$ and $-\chi_{0}\gamma_{l}$ correspond to $u_{1}$ and $u_{2}$ in Eqs.~(\ref{eq:2024-08-16-17-26}) and (\ref{eq:2024-08-16-17-27}). 
Hence, $(-\gamma_{l}\chi_{0}\Gamma_{l}\Lambda^{-1})$ and $(-\Lambda^{-1}\Gamma_{l}\chi_{0}\gamma_{l})$ are the products of $u_{1,2}$ type contributions and the constant coefficient $\Gamma_{l}\Lambda^{-1}$. 
Then, we employ a similar approximation to Eqs.~(\ref{eq:2024-08-16-17-26}) and (\ref{eq:2024-08-16-17-27}) also here. 
We employ the following approximation in ph channel ($l=\text{ph}$) 
\begin{align}
  (-\gamma_{l}\chi_{0}\Gamma_{l}\Lambda^{-1})_{\alpha\beta\gamma\lambda}(\omega,\nu) 
  &\approx C_{\alpha\gamma}(\omega)C_{\lambda\beta}(\omega+\nu), 
  \label{eq:2024-07-16-14-39}\\
  (-\Lambda^{-1}\Gamma_{l}\chi_{0}\gamma_{l})_{\alpha\beta\gamma\lambda}(\omega',\nu) 
  &\approx C_{\alpha\gamma}(\omega')C_{\lambda\beta}(\omega'+\nu). 
  \label{eq:2024-07-16-14-40}
\end{align} 
We can obtain this approximation for $\overline{\text{ph}}$ and pp channels by transforming the variables as $D\to C$ and $D \to P$ in Eq.~(\ref{eq:2020-05-10-14-49})-(\ref{eq:2020-05-10-14-53}).
Also, we introduce the following multiplication in terms of the correction factor $C$ for arbitrary vertex $V_{\alpha\beta\gamma\lambda}(\omega,\omega',\nu)$,
\begin{align}
  [C^{(2)}_{\text{ph}}V]_{\alpha\beta\gamma\lambda} =& \sum_{\alpha'\beta'}C_{\alpha\alpha'}(\omega)C_{\beta\beta'}(\omega+\nu)V_{\alpha'\beta'\gamma\lambda}, \label{eq:2024-08-19-13-18} \\
  [VC^{(2)}_{\text{ph}}]_{\alpha\beta\gamma\lambda} =& \sum_{\gamma'\lambda'}V_{\alpha\beta\gamma'\lambda'}C_{\gamma'\gamma}(\omega')C_{\lambda'\lambda}(\omega'+\nu), \label{eq:2024-08-19-13-19} \\
  [C^{(2)}_{\overline{\text{ph}}}V]_{\alpha\beta\gamma\lambda} =& \sum_{\alpha'\gamma'}C_{\alpha\alpha'}(\omega)V_{\alpha'\beta\gamma'\lambda}C_{\gamma'\gamma}(\omega'), \label{eq:2024-08-19-13-20} \\
  [VC^{(2)}_{\overline{\text{ph}}}]_{\alpha\beta\gamma\lambda} =& \sum_{\beta'\lambda'}C_{\beta\beta'}(\omega+\nu)V_{\alpha\beta'\gamma\lambda'}C_{\lambda'\lambda}(\omega'+\nu), \label{eq:2024-08-19-13-21} \\
  [C^{(2)}_{\text{pp}}V]_{\alpha\beta\gamma\lambda} =& \sum_{\alpha'\beta'}C_{\alpha\alpha'}(\omega)C_{\beta\beta'}(-\omega-\nu)V_{\alpha'\lambda\gamma\beta'}, \label{eq:2024-08-19-13-22} \\
  [VC^{(2)}_{\text{pp}}]_{\alpha\beta\gamma\lambda} =& \sum_{\gamma'\lambda'}V_{\alpha\lambda'\gamma'\beta} C_{\gamma'\gamma}(\omega')C_{\lambda'\lambda}(-\omega'-\nu), \label{eq:2024-08-19-13-23}
\end{align}
where we omit the frequency variables of $V$.
The diagrammatic expression of Eqs.~(\ref{eq:2024-08-19-13-18})-(\ref{eq:2024-08-19-13-23}) is shown in Fig.~\ref{fig:2024-05-26-15-48}~(a).
Then, we can rewrite Eq.~(\ref{eq:2024-05-19-14-46}) as  
\begin{align}
  \Phi_{l}
  =&
  C^{\text{(2)}}_{l} \Lambda 
  +  \Lambda C^{\text{(2)}}_{l}
  + C^{\text{(2)}}_{l} \Lambda C^{\text{(2)}}_{l}  \nonumber \\
  & + \Psi_{l}
  + C^{\text{(2)}}_{l} \Psi_{l}
  +  \Psi_{l}   C^{\text{(2)}}_{l}
  + C^{\text{(2)}}_{l} \Psi_{l}   C^{\text{(2)}}_{l}
  \nonumber \\
  &+ V_{l}^{\text{non-}C},
  \label{eq:2024-05-26-15-21}
\end{align} 
where, $ V_{l}^{\text{non-}C} = (\Gamma_{l}-\Lambda) [ \chi_{0} -  \chi_{0}(\Gamma_{l}\Lambda^{-1}-1) \Gamma_{l}\chi_{0} ] (\Gamma_{l}-\Lambda)$ is the vertex without the correction factor $C$.
By applying the approximation Eq.~(\ref{eq:2024-05-26-15-21}) for the reducible vertices, we can approximate the full-vertex as follows
\begin{align}
  F
  \approx&
  (\Lambda+\Psi_{\rm ph}) + C^{(2)}_{\text{ph}}(\Lambda+\Psi_{\rm ph}) \nonumber \\
  &+ (\Lambda+\Psi_{\rm ph})C^{(2)}_{\text{ph}} + C^{(2)}_{\text{ph}}(\Lambda+\Psi_{\rm ph})C^{(2)}_{\text{ph}}
  \nonumber \\
  &+ \Psi_{\overline{\text{ph}}} + C^{(2)}_{\overline{\text{ph}}}(\Lambda+\Psi_{\overline{\text{ph}}}) \nonumber \\
  &+ (\Lambda+\Psi_{\overline{\text{ph}}})C^{(2)}_{\overline{\text{ph}}} + C^{(2)}_{\overline{\text{ph}}}(\Lambda+\Psi_{\overline{\text{ph}}})C^{(2)}_{\overline{\text{ph}}}
  \nonumber \\
  &+ \Psi_{\text{pp}} + C^{(2)}_{\text{pp}}(\Lambda+\Psi_{\text{pp}}) \nonumber \\
  &+ (\Lambda+\Psi_{\text{pp}})C^{(2)}_{\text{pp}} + C^{(2)}_{\text{pp}}(\Lambda+\Psi_{\text{pp}})C^{(2)}_{\text{pp}}
  \nonumber \\
  &  + V^{\text{non-}C}_{\rm ph}   + V^{\text{non-}C}_{\overline{\text{ph}}}   + V^{\text{non-}C}_{\text{pp}}, 
  \label{eq:2023-09-26-02-16}
\end{align}
The diagrammatic expression of this simplified full-vertex is shown in Fig.~\ref{fig:2024-05-26-15-48}~(b).
To avoid the complexity of the equations and clarify the orbital and frequency dependences, 
we define the following expression
\begin{align}
  {\cal C}&[V(D)]_{D} \nonumber \\
  &= 
  V(D)
  + C^{(2)}_{\text{ph}}V(D) 
  + V(D)C^{(2)}_{\text{ph}}
  + C^{(2)}_{\text{ph}}V(D)C^{(2)}_{\text{ph}},
  \label{eq:2024-06-20-13-50} \\ 
  {\cal C}&[V(D)]_{C} \nonumber \\
  &= 
  V(C)
  + C^{(2)}_{\overline{\text{ph}}}V(C) 
  + V(C)C^{(2)}_{\overline{\text{ph}}}
  + C^{(2)}_{\overline{\text{ph}}}V(C)C^{(2)}_{\overline{\text{ph}}},
  \label{eq:2024-06-20-13-51} \\
  {\cal C}&[V(D)]_{P} \nonumber \\
  &= 
  V(P)
  + C^{(2)}_{\text{pp}}V(P) 
  + V(P)C^{(2)}_{\text{pp}}
  + C^{(2)}_{\text{pp}}V(P)C^{(2)}_{\text{pp}}.
  \label{eq:2024-06-20-13-52} 
\end{align}
Using Eqs.~(\ref{eq:2024-08-19-13-18})-(\ref{eq:2024-06-20-13-52}), we can express the simplified full-vertex as
\begin{align}
  F(D)
  \approx&
  {\cal C}[\Lambda + \Psi_{\text{ph}}]_{D}
  -
  {\cal C}[\Lambda + \Psi_{\text{ph}}]_{C}
  + 
  {\cal C}[\Lambda + \Psi_{\text{pp}}]_{P}
  \nonumber \\
  & + 
  \Lambda(C) 
  - 
  \Lambda(P) 
  \nonumber \\
  &+ V^{\text{non-}C}_{\rm ph}(D)   - V^{\text{non-}C}_{{\text{ph}}}(C)   + V^{\text{non-}C}_{\text{pp}}(P). 
  \label{eq:2024-06-23-16-24}
\end{align}

\subsection{Improved version of Self-energy to Full-vertex (S2F)}
In our previous study~\cite{doi:10.7566/JPSJ.91.034002}, we developed a method to estimate the two-particle full-vertex from the one-particle self-energy. 
We named this method "Self-energy to Full-vertex (S2F)". 
Here, we show the S2F procedure for the improved version of the simplified full-vertex described in the previous subsection.

In the improved version of the S2F procedure, we substitute the relation $\Gamma_{l} = \Lambda + \gamma_{l}$ to only the left $\Gamma_{l}$ of $\chi_{l}$ in the reducible vertex $\Phi_{l} = -\Gamma_{l}\chi_{l}\Gamma_{l}$, not both of $\Gamma_{l}$ as in Eq.~(\ref{eq:2024-03-22-00-52}),
\begin{align}
  \Phi_{l}
  =&
  -\Gamma_{l}\chi_{l}\Gamma_{l}
  \nonumber \\
  =&
  -(\Lambda+\gamma_{l}) \chi_{l} \Gamma_{l}
  \nonumber \\
  =&
  -\Lambda\chi_{l}\Gamma_{l} -\gamma_{l}(\chi_{0}-\chi_{0}\Gamma_{l}\chi_{l})\Gamma_{l}
  \nonumber \\
  =&
  -\Lambda\chi_{l}\Gamma_{l} + (-\gamma_{l}\chi_{0}\Gamma_{l}\Lambda^{-1})\Lambda + (-\gamma_{l}\chi_{0}\Gamma_{l}\Lambda^{-1})\Lambda\chi_{l}\Gamma_{l}
  \nonumber \\
  =&
  C^{(2)}_{l}\Lambda + (1+C^{(2)}_{l}) \tilde{\Psi}_{l},
  \label{eq:2024-05-26-16-47}
\end{align}
where $\tilde{\Psi}_{l}=-\Lambda\chi_{l}\Gamma_{l}$.
We define the local irreducible susceptibilities in ph and pp channels as 
\begin{align}
  \chi_{0}(\nu) =& -\sum_{n}G(\omega)G(\omega+\nu), \\
  \phi_{0}(\nu) =& \sum_{n}G(\omega)G(-\omega-\nu),
\end{align}
and the Green's function like quantities as 
\begin{align}
  G^{\rm CL}(\omega) =& C(\omega)G(\omega),  \label{eq:2024-08-20-14-20} \\
  G^{\rm CR}(\omega) =& G(\omega)C(\omega), \label{eq:2024-08-20-14-21} \\
  G^{\rm CLR}(\omega) =& C(\omega)G(\omega)C(\omega). \label{eq:2024-08-20-14-22}
\end{align}
Now, we can obtain the approximation form of the correlation part of the self-energy (the self-energy subtracted by the static Hatree-Fock terms) as follows by using Eqs.~(\ref{eq:2020-05-09-22-20}) and (\ref{eq:2024-05-26-16-47})-(\ref{eq:2024-08-20-14-22}).
\begin{align}
  \Sigma(\omega)
  =&
  \dfrac{1}{2}\sum F(D)\chi_{0}(D)\Lambda(D) \nonumber \\
  \approx&
  Y(\omega) - C(\omega)X(\omega)
\end{align}
where, 
\begin{align}
  Y_{\alpha\beta}(\omega)
  =&
  \dfrac{1}{2}
  \sum_{\nu} [\Lambda(D) \chi_{0}(\nu)\Lambda(D)]_{\alpha\gamma\beta\lambda}G_{\beta\lambda}(\omega+\nu)
  \nonumber \\
  &+
  \sum_{m} [\tilde{\Psi}_{\rm ph}(\nu)\chi_{0}(\nu)\Lambda(D)]_{\alpha\gamma\beta\lambda}G_{\beta\lambda}(\omega+\nu)
  \nonumber \\
  &-
  \dfrac{1}{2}\sum_{\nu} [\tilde{\Psi}_{\rm pp}(\nu)\phi_{0}(\nu)\Lambda(D)]_{\alpha\gamma\beta\lambda}G_{\gamma\lambda}(-\omega-\nu),
  \label{eq:2024-05-26-16-52} \\
  X_{\alpha\beta}(\omega)
  =&
  \sum_{\nu} [\Lambda(D)\chi_{0}(\nu)\Lambda(D)]_{\alpha\gamma\beta\lambda}G_{\gamma\lambda}^{\rm CL}(\omega+\nu)
  \nonumber \\
  &+
  \sum_{m}[\tilde{\Psi}_{\rm ph}(\nu)\chi_{0}(\nu)\Lambda(D)]_{\alpha\gamma\beta\lambda}G_{\gamma\lambda}^{\rm CL}(\omega+\nu)
  \nonumber \\
  &+
  \dfrac{1}{2}\sum_{\nu}[\Lambda(P)\phi_{0}(\nu)\Lambda(P)]_{\alpha\gamma\beta\lambda}G_{\gamma\lambda}^{\rm CL}(-\omega-\nu)
  \nonumber \\
  &+
  \dfrac{1}{2}\sum_{\nu}[\tilde{\Psi}_{\rm pp}(\nu)\phi_{0}(\nu)\Lambda(P)]_{\alpha\gamma\beta\lambda}G_{\gamma\lambda}^{\rm CL}(-\omega-\nu).
  \label{eq:2024-05-26-16-53}
\end{align}
When we assume that the self-energy has already been obtained, we can estimate the correction factor $C$ in the following steps:
\begin{enumerate}
  \item[(i)] calculate $\tilde{\Psi}_{l}$ by the simplified parquet method~\cite{doi:10.1143/JPSJ.79.094707,PhysRevB.104.035160,doi:10.7566/JPSJ.91.034002}.
  \item[(ii)] calculate $X$ and $Y$ by Eqs.~(\ref{eq:2024-05-26-16-52}) and (\ref{eq:2024-05-26-16-53}).
  \item[(iii)] obtain $C$ by $C(\omega)=[Y(\omega)-\Sigma(\omega) ]X(\omega)^{-1}$.
  \item[(iv)] go back to (ii) (iterate until convergence). 
\end{enumerate}
After convergence, we can obtain the simplified form of the full-vertex by Eq.~(\ref{eq:2024-06-23-16-24}).

\section{Results} \label{sec:2024-07-07-15-52}

\begin{figure*}[t]
  \centering
  {\includegraphics[width=180mm,clip]{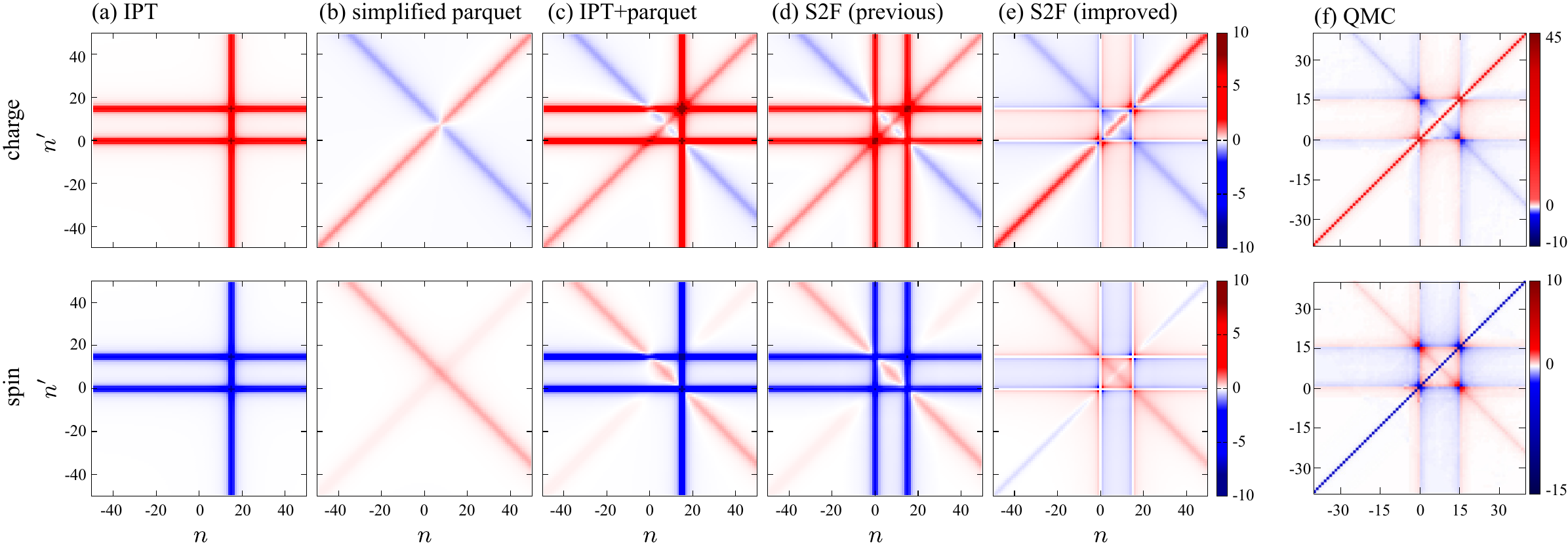}} 
  \caption{
    Comparison of the full-vertices obtained by 
    (a) IPT, (b) simplified parquet, (c) IPT+parquet, (d) previous S2F, (e) improved S2F, and (f) QMC.
    The calculation is performed for the cubic lattice model. 
    The interaction strength, the temperature, and the bosonic frequency are $U/D=2$, $T/D=1/8$, and $\nu_{m}=-30\pi T \ (m=-15)$, respectively.
    The figures in (f) adapted with permission from Ref.~\cite{PhysRevB.96.035114}. Copyrighted by the American Physical Society.
  }
  \label{fig:2024-06-27-16-22}
\end{figure*}

We here compare the improved S2F results with the ones obtained by the numerically exact CT-QMC and ED methods~\cite{PhysRevB.96.035114,PhysRevB.86.125114} and the methods that can give the simplified full-vertex with inexpensive numerical costs: the iterative perturbation theory~(IPT)~\cite{doi:10.1143/PTPS.46.244, doi:10.1143/PTP.53.970, doi:10.1143/PTP.53.1286, Yamada4, PhysRevB.45.6479}, the simplified parquet~\cite{doi:10.1143/JPSJ.79.094707,PhysRevB.104.035160,doi:10.7566/JPSJ.91.034002}, IPT+parquet~\cite{PhysRevB.104.035160}, and the previous S2F~\cite{doi:10.7566/JPSJ.91.034002}. 
In the S2F procedures, we use IPT+parquet~\cite{PhysRevB.104.035160} to obtain the input self-energy. 
We consider that IPT+parquet has enough accuracy for the comparison in this study since the IPT+parquet can give close results to the numerically exact CT-QMC~[see Ref.~\cite{PhysRevB.104.035160} for detail].
We employ the single-orbital cubic lattice model with only the nearest neighbor hopping at half-filling.
The system we consider here has the $SU(2)$ symmetry in the spin space, and so we can divide the full-vertex into four channels $c$~(charge), $s$~(spin), $e$~(even), and $o$~(odd) in terms of the parity of spin as follows
\begin{align}
  F_{c}(D) =& F_{\uparrow\uparrow\uparrow\uparrow}(D) + F_{\uparrow\uparrow\downarrow\downarrow}(D) \label{eq:2024-11-12-16-41}, \\
  F_{s}(D) =& F_{\uparrow\uparrow\uparrow\uparrow}(D) - F_{\uparrow\uparrow\downarrow\downarrow}(D) \label{eq:2024-11-12-16-42}, \\
  F_{e}(P) =& F_{\uparrow\downarrow\uparrow\downarrow}(P) - F_{\uparrow\downarrow\downarrow\uparrow}(P) \label{eq:2024-11-12-16-43}, \\
  F_{o}(P) =& F_{\uparrow\downarrow\uparrow\downarrow}(P) + F_{\uparrow\downarrow\downarrow\uparrow}(P) \label{eq:2024-11-12-16-44}. 
\end{align}

Fig.~\ref{fig:2024-06-27-16-22} shows the simplified full-vertices in the charge and spin channels obtained by (a)~IPT, (b)~simplified parquet, (c)~IPT+parquet, (d)~previous S2F, (e)~improved S2F, and the numerically exact full-vertices obtained by (f)~CT-QMC~\cite{PhysRevB.96.035114}.
The bare-vertex is always subtracted from the full-vertex in this paper.
The interaction strength and the temperature are $U/D=2$ and $T/D=1/8$, respectively, which are the same as ones used in Fig.~5 in Ref.~\cite{PhysRevB.96.035114}.
$D/2=\sqrt{6}t$ is the standard deviation of the density of states (DOS) of the cubic lattice with only the nearest neighbor hopping.
To show the systematic improvements of the simplified vertex and clarify the roles of ingredients of the full-vertex, we explain the results of the methods (a)-(e) in order below.

In our previous study~\cite{PhysRevB.104.035160}, we showed that we can interpret IPT as an approximation using the simplified full-vertex, although it is generally regarded as an interpolation method based on the second-order perturbation self-energy.
The IPT full-vertex has the following form,
\begin{align} 
  F(\omega&,\omega',\nu) \nonumber \\
  \approx&
  C_{2}(\omega)C_{1}(\omega+\nu)U
  C_{1}(\omega')C_{1}(\omega'+\nu), 
  \label{eq:2024-07-03-16-24} \\
  C_{1} &= G_{0}G^{-1}, 
  \label{eq:2024-07-03-16-26} \\
  C_{2} &= [I-B\Sigma^{\text{2nd}}]^{-1}A,
  \label{eq:2024-07-03-16-27}
\end{align}
where $G_{0}$ and $G$ are non-interacting and interacting Green's functions.
$\Sigma^{\text{2nd}}$ is the second-order perturbation self-energy.
$A$ and $B$ are the parameters that we determine interpolatively. 
We can see from Fig.~\ref{fig:2024-06-27-16-22} (a) and Eq.~(\ref{eq:2024-07-03-16-24}) that the IPT full-vertex has the frequency dependence that gives the cross and central structures,
although their behaviors are not precisely consistent with the exact one~(we mention this point below).
In our interpretation, since IPT full-vertex has the cross and central structures, IPT can capture the strong correlation effects [see Ref.~\cite{PhysRevB.104.035160} for detail]. 
However, IPT fails the diagonal structure and violates the symmetry for the exchange of $\omega$ and $\omega'$ coming from the crossing symmetry. 

In contrast, the full-vertex obtained by the simplified parquet method has only the diagonal structure as shown in Fig.~\ref{fig:2024-06-27-16-22}~(b). 
This is because the simplified parquet we employ here~\cite{doi:10.1143/JPSJ.79.094707,PhysRevB.104.035160,doi:10.7566/JPSJ.91.034002} reduces its numerical cost by dropping the fermionic frequency of the two-particle susceptibilities. 
The full-vertex in the simplified parquet takes the following form,
\begin{align}
  F&(\omega,\omega',\nu) \nonumber \\
  \approx&
  \Lambda + \Phi_{\rm ph}(\nu) 
  + \Phi_{\rm \overline{ph}}(\omega-\omega') + \Phi_{\rm pp}(\omega+\omega'+\nu).
  \label{eq:2024-07-03-17-02}
\end{align}
Only the bosonic frequency dependence that gives the diagonal structures emerges in Eq.~(\ref{eq:2024-07-03-17-02}).

As mentioned above, IPT can capture the strong correlation effect by the cross and central structures. 
However, IPT cannot give good results in multi-band systems, especially in non-degenerate systems.
We have overcome the drawback of IPT by combining IPT and the simplified parquet method (IPT+parquet)~\cite{PhysRevB.104.035160}.
The full-vertex in IPT+parquet has the following form,
\begin{align} 
  F(\omega&,\omega',\nu) \nonumber \\
  \approx&
  C_{2}(\omega)C_{1}(\omega+\nu)
  F_{0}(\omega,\omega',\nu)
  C_{1}(\omega')C_{1}(\omega'+\nu), 
  \label{eq:2024-07-04-15-01} \\
  F_{0}(\omega&,\omega',\nu) \nonumber \\
  \approx&
  \Lambda + \Phi_{\rm ph}(\nu) 
  + \Phi_{\rm \overline{ph}}(\omega-\omega') + \Phi_{\rm pp}(\omega+\omega'+\nu),
  \label{eq:2024-07-04-15-04} \\
  C_{1} &= G_{0}G^{-1}, 
  \label{eq:2024-07-04-15-02} \\
  C_{2} &= [I-B\Sigma^{\text{cr}}]^{-1}A,
  \label{eq:2024-07-04-15-03}
\end{align}
where $\Sigma^{\text{cr}}$ is the correlation part of the self-energy.
We can see that the IPT+paruqet full-vertex in Eqs.~(\ref{eq:2024-07-04-15-01}), (\ref{eq:2024-07-04-15-04}) and Fig.~\ref{fig:2024-06-27-16-22}~(c) has the frequency dependence that gives the diagonal, cross, and central structures. 

As mentioned above, the IPT full-vertex violates the symmetry for exchanging $\omega$ and $\omega'$. 
This is also true for IPT+parquet.
These drawbacks come from the fact that the correction factors in IPT solution in Eqs.(\ref{eq:2024-07-03-16-26}) and (\ref{eq:2024-07-03-16-27}) and IPT+parquet solution in Eqs.~(\ref{eq:2024-07-04-15-02}) and (\ref{eq:2024-07-04-15-03}) do not satisfy the condition $C_{1}=C_{2}$ that comes from the crossing symmetry of the full-vertex.
Then, we have developed the previous S2F to obtain the simplified full-vertex while satisfying the crossing symmetry. 
The full-vertex in the previous S2F can be written as in Eqs.~(\ref{eq:2021-04-14-21-16}) and (\ref{eq:2021-04-14-21-17}). 
The condition $C_{1}=C_{2}\equiv \tilde{C}$ is satisfied, and therefore the cross structures in the full-vertex obtained by the previous S2F are symmetric for the exchange of $\omega$ and $\omega'$ as shown in Fig.~\ref{fig:2024-06-27-16-22}~(d). 
Although the full-vertex obtained by the previous S2F roughly reproduces the exact full-vertex obtained by CT-QMC~(Fig.~\ref{fig:2024-06-27-16-22}~(f)), there are some qualitatively inconsistent behaviors between the two full-vertices. 
For example, the full-vertex obtained by the previous S2F has peak structures at $\omega^{(\prime)}=0$ and $\omega^{(\prime)}+\nu=0$ lines, while the exact full-vertex obtained by CT-QMC changes its value sharply at these lines.
Also, the signs of their diagonal structures sloping upward in the spin channel are opposite.

As shown in Fig.~\ref{fig:2024-06-27-16-22}~(e),
by using the improved S2F, we can obtain the full-vertex that sharply changes its value at $\omega^{(\prime)}=0$ and $\omega^{(\prime)}+\nu=0$ lines, and the diagonal structure sloping upward in the spin channel has the same sign with the exact full-vertex.
We achieve the improvement from the previous S2F because we refine the formulation of the simplification of the full-vertex to be more faithful to the diagrammatic structure.
We explain this point in detail below. 

\begin{figure}
  \centering
  {\includegraphics[width=85mm,clip]{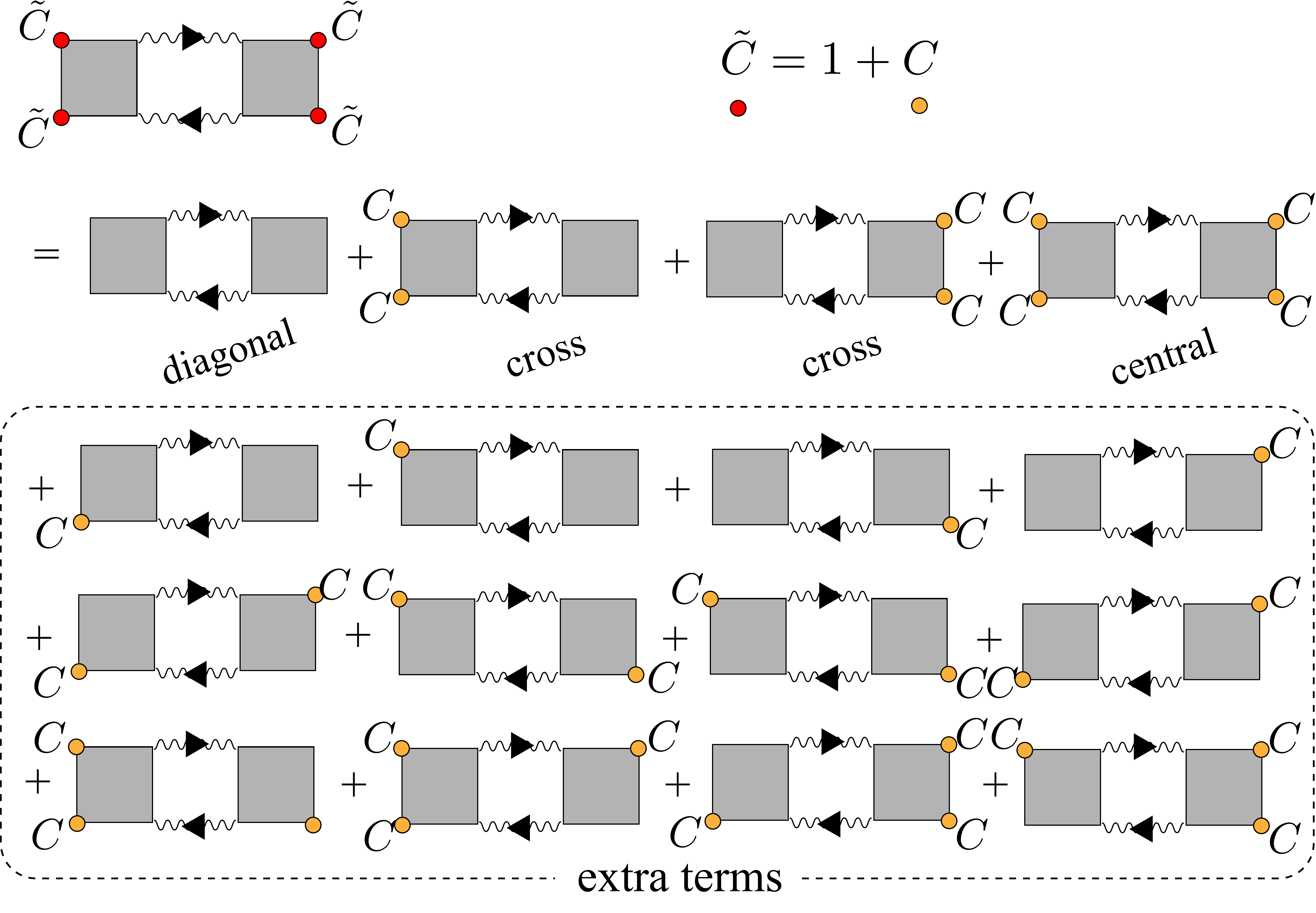}} 
  \caption{  
    The decomposition of the previous version of a simplified vertex.
    The simplified vertex has extra terms odd in terms of $C$ that cannot appear in a diagrammatic manner.
  }
  \label{fig:2024-10-23-13-56}
\end{figure}

When the previous version of the simplified full-vertex in Eqs.~(\ref{eq:2021-04-14-21-16}) and (\ref{eq:2021-04-14-21-17}) can have the three characteristic structures (diagonal, cross, and cental), the correction factor $\tilde{C}$ can be divided into unity and frequency-dependent terms as 
\begin{align}
  \tilde{C}(\omega) = 1 + C(\omega).
  \label{eq:2024-10-23-13-34}
\end{align}
We use the letter ``$C$" as the frequency-dependent term since it has the same roles as the correction factor $C$ in the improved version of the simplified full-vertex in Eq.~(\ref{eq:2024-06-23-16-24}).
We can see that the previous version of the simplified full-vertex has not only the terms that give the three characteristic structures but also extra terms odd in terms of $C$ by substituting Eq.~(\ref{eq:2024-10-23-13-34}) to Eqs.~(\ref{eq:2021-04-14-21-16}) and (\ref{eq:2021-04-14-21-17}).
The diagrammatic expression of the decomposition of the previous version of a simplified vertex is shown in Fig.~\ref{fig:2024-10-23-13-56}.
Since the $C$ terms have the role of approximating the frequency behavior that mainly comes from the pair of the Green's functions that connect the two (b)-type vertices as in Eqs.~(\ref{eq:2024-08-16-17-26}) and (\ref{eq:2024-08-16-17-27}) in Sect.~\ref{sec:2024-07-01-14-44} and in Eqs.~(\ref{eq:2024-07-16-14-39}) and (\ref{eq:2024-07-16-14-40}) in Sect.~\ref{sec:2024-10-23-14-21}, the correction factor $C$ must appear in pairs and be expected to have the Green's function-like frequency behavior.

\begin{figure}[t]
  \centering
  {\includegraphics[width=85mm,clip]{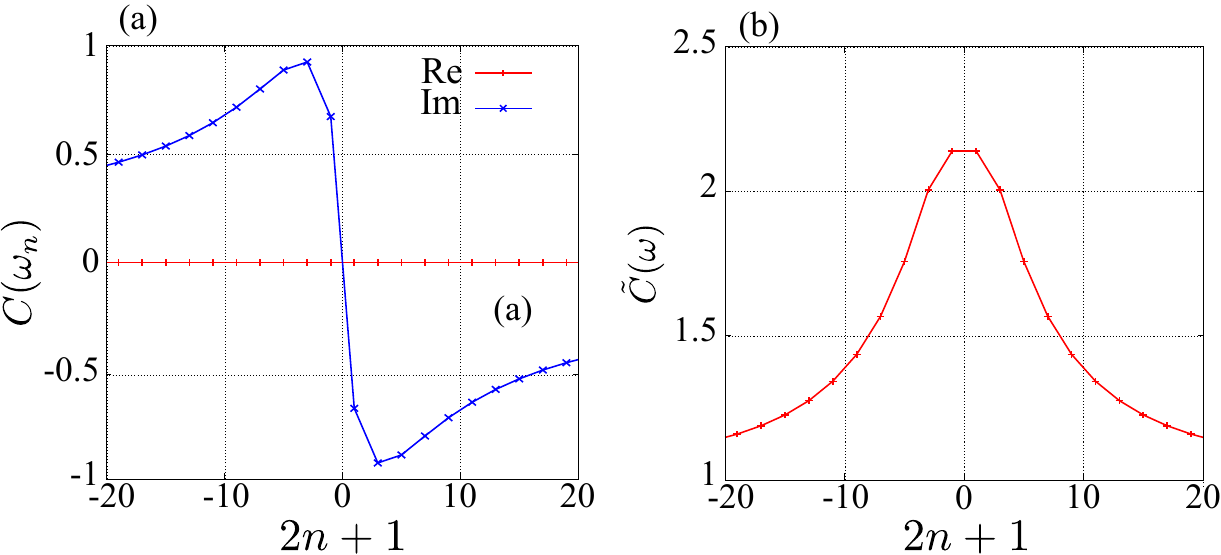}} 
  \caption{  
    The correction factor $C(\omega)$ for the cubic lattice model obtained by (a)~the improved S2F and (b)~the previous S2F.
    The interaction strength and the temperature are $U/D=2$ and $T/D=1/26$, respectively.
    (a) The solid line with $+$ and $\times$ symbols represent the real and imaginary parts, respectively. 
    (b) Only the real part is shown since the imaginary part is zero in this situation.
  }
  \label{fig:2024-06-25-13-42}
\end{figure}

In the improved version of the simplification, we construct the simplified full-vertex not to have these extra terms by treating the three characteristic structures separately, as shown in Fig.~\ref{fig:2024-10-15-16-00}.
This can make the correction factor $C(\omega)$ have the Green-function-like frequency behavior as in Fig.~\ref{fig:2024-06-25-13-42}~(a).
As mentioned in Sect.~\ref{sec:2024-07-01-14-44}, the frequency dependence of the cross and central structures reflects that of the Green's functions that connect the two vertices in different channels.
And in this study, since we consider the electron-hole symmetric case, Green's function is the pure imaginary function.
The full-vertex sharply changes its value at $\omega^{(\prime)}=0$ and $\omega^{(\prime)}+\nu=0$ lines, reflecting the nature of the ${\rm Im}G(\omega)$.
We can reproduce this behavior by the correction factor $C(\omega)$ that has the Green's function-like frequency dependence as in Fig.~\ref{fig:2024-06-25-13-42}~(a).
On the other hand, the correction factor $\tilde{C}(\omega)$ of the previous S2F~\cite{doi:10.7566/JPSJ.91.034002} in Eq.~(\ref{eq:2021-04-14-21-16}) has the Lorentzian like frequency behavior as in Fig.~\ref{fig:2024-06-25-13-42}~(b). 
Therefore, the full-vertex obtained by the previous S2F just has peaks at $\omega^{(\prime)}=0$ and $\omega^{(\prime)}+\nu=0$ lines as shown in Fig.~\ref{fig:2024-06-27-16-22}~(d).

The refinement of the formulation also makes the diagonal structure closer to the exact CT-QMC results than the previous S2F. 
The terms that give diagonal structures in the full-vertex in the charge channel are
\begin{align}
  \Psi_{c}(\nu) 
  -\dfrac{1}{2}[\Psi_{c}+3\Psi_{s}](\omega-\omega')
  + [\Psi_{e}-3\Psi_{o}](\omega+\omega'+\nu) 
  \label{eq:2024-07-01-14-01}
\end{align}
and in the spin channel
\begin{align}
  \Psi_{s}(\nu) 
  -\dfrac{1}{2}[\Psi_{c} - \Psi_{s}](\omega-\omega') 
  -[\Psi_{e}-\Psi_{o}](\omega+\omega+\nu),
  \label{eq:2024-07-01-14-02}
\end{align}
where $\Psi_{r}=-\Lambda_{r}\chi_{r}\Lambda_{r}$ and $r=(c,s,e,o)$ represent the charge, spin, even, and odd channels.
Eqs.~(\ref{eq:2024-07-01-14-01}) and (\ref{eq:2024-07-01-14-02}) correspond to a part of Eqs.(25) and (26) in our previous S2F paper~\cite{doi:10.7566/JPSJ.91.034002}, where $\Phi_{r}=-\Gamma_{r}\chi_{r}\Gamma_{r}$ give the diagonal structures.
The replacement of the irreducible vertex $\Gamma_{r}$ with the fully-irreducible vertex $\Lambda_{r}$ is derived from the refinement of the simplification method of the full-vertex. 
This replacement of the vertices reduces the overestimation (underestimation) of the charge (spin) fluctuation in the simplified parquet method.
As a result, the strength of the diagonal structures in the improved S2F is larger and closer to the exact results~\cite{PhysRevB.96.035114} than the previous S2F~\cite{doi:10.7566/JPSJ.91.034002}.

\begin{figure}[t]
  \centering
  {\includegraphics[width=85mm,clip]{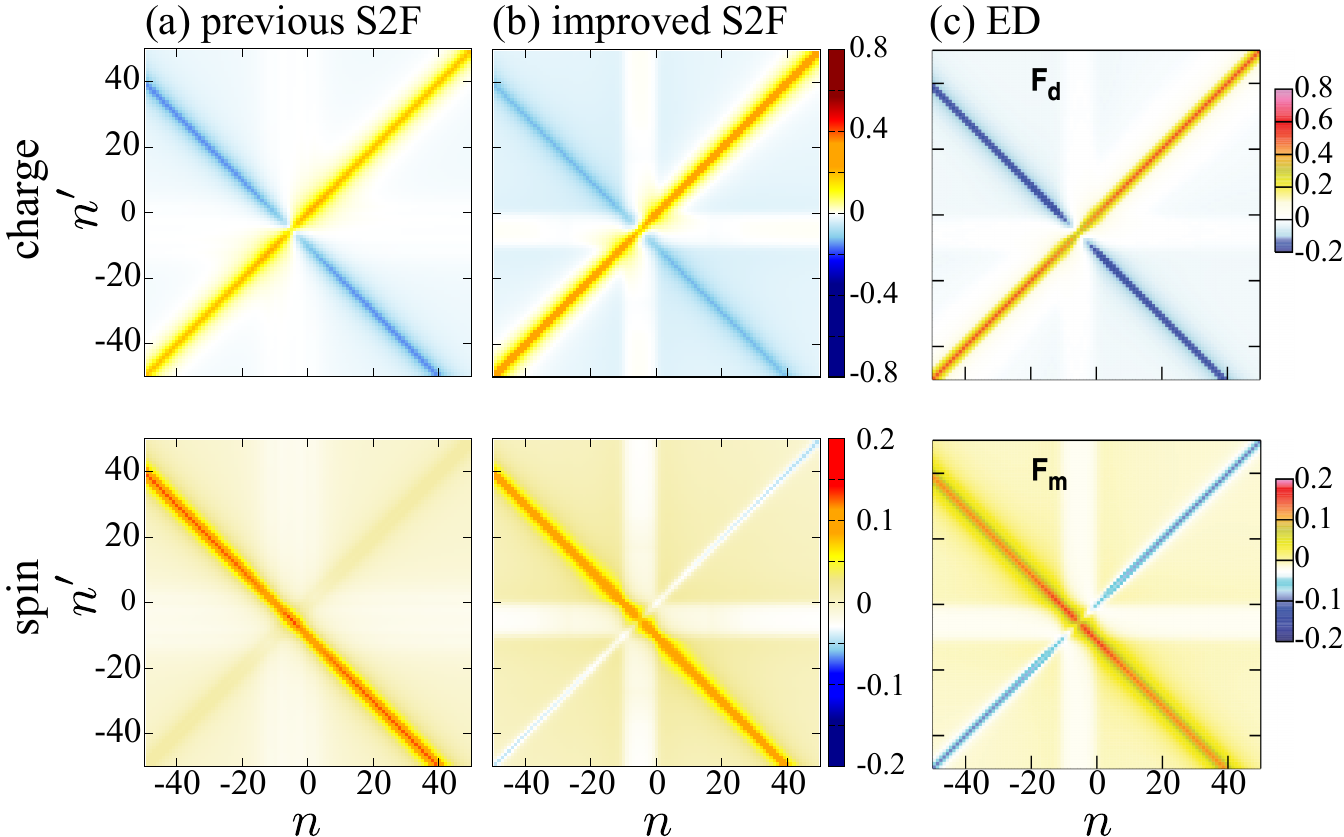}} 
  \caption{  
    The full-vertex for the cubic lattice model at $\nu=20 \pi T \ (m=10)$ obtained by (a) the previous S2F, (b) the improved S2F, and (c) ED, where the bare vertex is subtracted in each figure.
    The upper (lower) panel shows the charge (spin) channel. 
    The interaction strength and the temperature are $U/D=0.5$ and $T/D=1/26$, respectively. 
    $D/2=\sqrt{6}t$ is the standard deviation.
    The figures in (c) adapted with permission from Ref.~\cite{PhysRevB.86.125114}. Copyrighted by the American Physical Society.
  }
  \label{fig:2024-06-12-16-19}
\end{figure}

\begin{figure}[t]
  \centering
  {\includegraphics[width=85mm,clip]{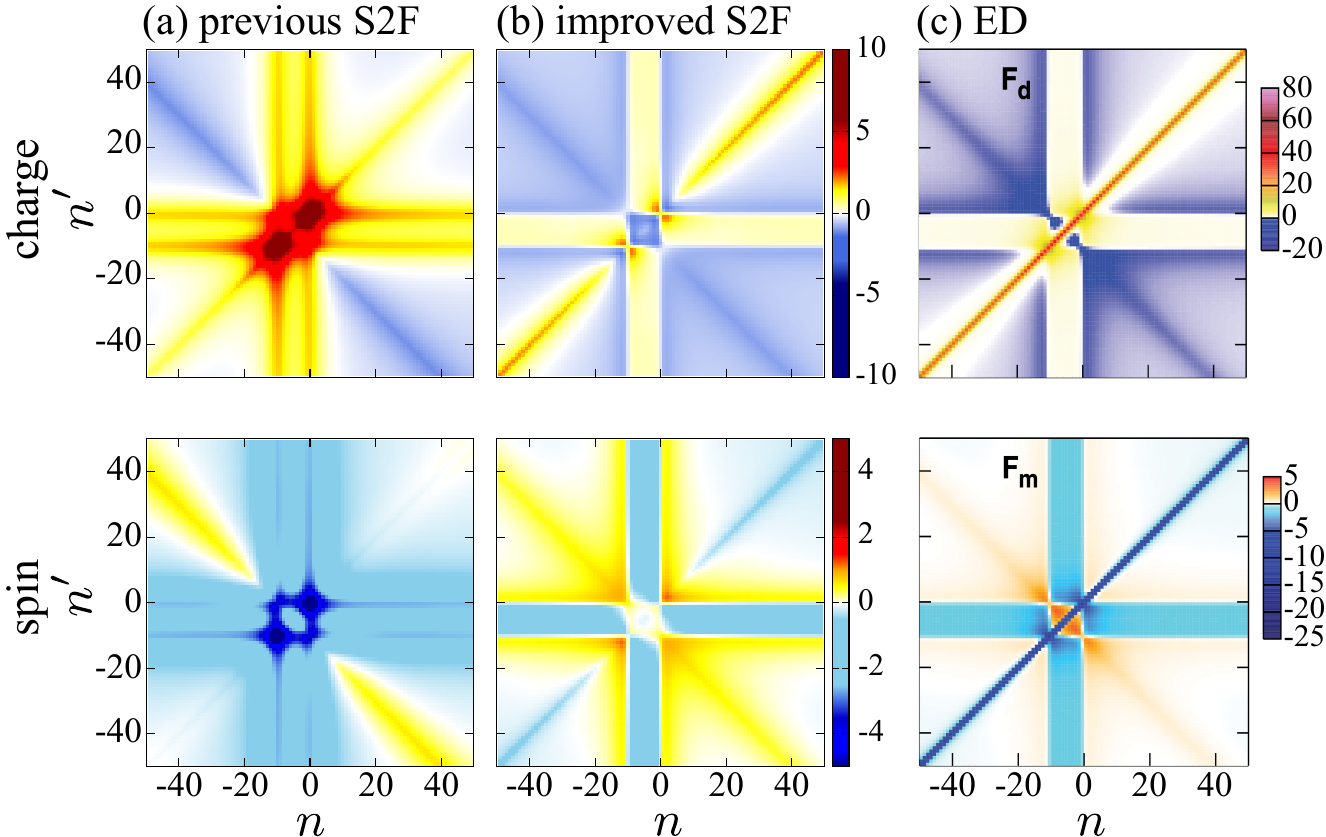}} 
  \caption{  
    The full-vertex for the cubic lattice model at $\nu=20 \pi T \ (m=10)$ obtained by (a) the previous S2F, (b) the improved S2F, and (c) ED, where the bare vertex is subtracted in each figure.
    The upper (lower) panel shows the charge (spin) channel. 
    The interaction strength and the temperature are $U/D=2$ and $T/D=1/26$, respectively. 
    $D/2=\sqrt{6}t$ is the standard deviation.
    The figures in (c) adapted with permission from Ref.~\cite{PhysRevB.86.125114}. Copyrighted by the American Physical Society.
  }
  \label{fig:2024-06-12-16-20}
\end{figure}

Next, we show the results at a lower temperature, comparing the improved S2F with the previous S2F and the DMFT+ED\cite{PhysRevB.86.125114}.
Figures~\ref{fig:2024-06-12-16-19} and \ref{fig:2024-06-12-16-20} show the full-vertex of the cubic lattice model in the charge and spin channels obtained by the previous S2F, improved S2F, and ED.
In Fig.~\ref{fig:2024-06-12-16-19},  the interaction strength, the temperature, and the bosonic frequency are $U/D=0.5$, $T/D=1/26$, and $\nu=20\pi T \ (m=10)$, respectively, which are the same as the ones used in Fig.~7 in Ref.~\cite{PhysRevB.86.125114}.
We can see that the improved S2F shows the improvement on the points explained above from the previous S2F, namely the strength of the diagonal structure and the sharpness of the cross structure at $\omega^{(\prime)}=0$ and $\omega^{(\prime)}+\nu=0$ lines.
In Fig.~\ref{fig:2024-06-12-16-20},  the interaction strength, the temperature, and the bosonic frequency are $U/D=2$, $T/D=1/26$, and $\nu=20\pi T \ (m=10)$, respectively, which are the same as the ones used in Fig.~9 in Ref.~\cite{PhysRevB.86.125114}.
The full-vertex obtained by the improved S2F changes its value sharply at $\omega^{(\prime)}= 0$ and $\omega^{(\prime)}+\nu=0$ lines, consistent with ED results~\cite{PhysRevB.86.125114}, while the full-vertex obtained by the previous S2F has broad peak structures at these lines.
The strength of the diagonal structure of the improved S2F is closer to the numerically exact ED result than the previous S2F also here.

\section{Discussion: Comparison to other methods} \label{sec:2024-07-11-13-14}

Let us mention here the advantages of our method developed in this study.
In the S2F method, we can construct the two-particle full-vertex from the one-particle self-energy, which means that any impurity solver that gives the one-particle self-energy can be plugged into our method. 
In addition, efficient computational cost is a notable advantage of the S2F method.
Let us compare the computational cost for calculating the full-vertex with S2F and numerically exact CT-QMC.
We show how the computational time scales when estimating the local full-vertex in the first column in Table~\ref{tab:2025-02-23-16-59}.
We also show how the computational time of IPT and the simplified parquet scales for comparison. 
Although several versions of CT-QMC~\cite{PhysRevB.72.035122, Rubtsov2004,PhysRevB.80.195111,PhysRevB.83.075122,PhysRevB.89.195146,Seth2016274,PhysRevLett.97.076405,PhysRevB.74.155107,Gull_phdthesis,lewin_thesis,triqs_ctqmc_solver_legendre} exist, we here assume the hybridization expansion (CT-HYB)~\cite{Seth2016274,PhysRevLett.97.076405,PhysRevB.74.155107,Gull_phdthesis,lewin_thesis,triqs_ctqmc_solver_legendre}, which is efficient in the strongly correlated and low-temperature regime.
For example, the core hours ($=$ the number of CPU cores $\times$ the number of hours for the calculation) of S2F is about an order less than that of CT-QMC when estimating the local full-vertex in the single-band case. 
As shown in Table~\ref{tab:2025-02-23-16-59}, the computational time of CT-QMC for obtaining the local full-vertex scales as $N_{\omega}^{3}N_{o}^{4}\times 2^{\alpha N_{o}}$.
$N_{\omega}$, $N_{o}$ are the number of Matsubara frequencies and orbitals. 
The factor $\alpha$ depends on the optimization methods~\cite{PhysRevB.80.235117,PhysRevB.90.075149,Shinaoka2017} for calculating the trace in the impurity Hilbert space~($\alpha=6$ with no optimization). 
The computational time of S2F, IPT, and the simplified parquet depends on the situation. 
When the number of orbitals is small, their computational time scales as $N_{\omega}\log(N_{\omega})N_{o}^{4}$ since the Fourier transformation is the most expensive part. 
When the number of orbitals is large, their computational time scales as $N_{\omega}N_{o}^{6}$ since the multiplication of two-particle quantities is the most expensive part.
Therefore, the computational time of CT-QMC increases much faster than that of S2F, IPT, and the simplified parquet. 

Using the simplified form of the full-vertex,
we can develop numerically very inexpensive diagrammatic extensions of DMFT. 
We show how the computational time for executing the dual fermion calculation, one of the diagrammatic extensions, scales in the second column in Table~\ref{tab:2025-02-23-16-59}.
In our previous study~\cite{doi:10.7566/JPSJ.91.034002}, we developed the efficient dual fermion method~(EDF), where we treat only one frequency variable in practical calculation even though the full-vertex has three frequency variables [see Ref.~\cite{doi:10.7566/JPSJ.91.034002} for details]. 
The computational time of EDF depends on the situation for the same reason as in the S2F, IPT, and simplified parquet cases.
Namely, when $N_{o}$ is small, it scales as $N_{\omega}N_{k}\log(N_{\omega}N_{k})N_{o}^{4}$ since the Fourier transformation is the most expensive part.
On the other hand, when $N_{o}$ is large, it scales as $N_{\omega}N_{k}N_{o}^{6}$ since the multiplication of two-particle quantities is the most expensive part.
The computational time of the ordinary dual fermion method scales as $(N_{\omega}N_{k})^{4}N_{o}^{6}$. 
The computational time of the ordinary dual fermion method increases $(N_{\omega}N_{k})^{3}N_{o}^{2}/\log(N_{\omega}N_{k})$ or $(N_{\omega}N_{k})^{3}$ times faster than EDF.  
The full-vertex obtained by IPT and the simplified parquet can be used in the EDF.
The simplified full-vertex improved in this study can also be used in the EDF, although the procedures become slightly complex.
Combining the improved version of the simplified full-vertex with EDF and developing other numerically efficient diagrammatic extensions of DMFT are future works.

Given the above, let us discuss situations where these methods are suitable or not as shown in Table~\ref{tab:2025-02-23-16-59}.
First, we consider the single orbital case.
 CT-QMC is suitable for precise analysis since it is numerically exact.
However, S2F, IPT, and the simplified parquet are suitable at low-temperatures since their computational time increases linearly with the number of Matubara frequencies while that of CT-QMC increases cubically.
On the other hand, at present, we find that we cannot address the strong charge fluctuation in S2F and the simplified parquet.
The simplified parquet tends to underestimate (overestimate) the spin (charge) fluctuation. 
When the charge fluctuation becomes large, it cannot stop growing because of insufficient negative feedback from the spin channel. Then, the self-consistent loop becomes unstable. 
Also, IPT cannot analyze systems with large two-body fluctuation since IPT full-vertex has no diagonal structure, which is crucial for the two-particle correlation, while it has cross and central structures.
In contrast, the full-vertex of simplified parquet has no cross and central structure, while it has diagonal structures, so we cannot address the strong correlation regime using the simplified parquet.

When we consider the multi-orbital case, S2F, IPT, and the simplified parquet have a significant advantage in terms of numerical costs.
Especially in S2F, we can obtain the local full-vertex with much lower numerical cost than CT-QMC, while keeping the essential frequency structures.
On the other hand, we do not have a reasonable basis for decoupling the orbital variables in the approximation made in Eqs.~(\ref{eq:2024-07-16-14-39}) and (\ref{eq:2024-07-16-14-40}), while the decoupling of the frequency variables is based on the diagrammatic structures and the nature of the vertex functions as explained in Sect.~\ref{sec:2024-10-23-14-21}.
Therefore, for example in systems with strong orbital hybridization, S2F might break down.
Examining the validity of the decoupling of orbital variables is an important future work.
Another drawback in S2F is that S2F cannot be applied to the negative Hund coupling regime since the parquet loop for obtaining diagonal structures becomes unstable.
This is because the charge susceptibility cannot stop growing due to insufficient negative feedback from the spin channel to the charge channel. 
This drawback is common with the simplified parquet.

Finally, let us mention the possibility of further developments of approximations using our simplified full-vertex and other approaches that can support the validity of our simplification of the full-vertex. 
The simplified form of the full-vertex in Eq.~(\ref{eq:2024-06-23-16-24}) provides a hierarchy of approximations of the full-vertex.
S2F and EDF mentioned above are examples of the approximations in the hierarchy.
In fact, 
we can employ other approaches instead of the simplified parquet to obtain the vertices $\Psi_{l}$ that mainly convey the bosonic frequency in Eq.~(\ref{eq:2024-06-23-16-24}). 
Note that the hierarchy of approximations contains the exact solution.
Namely, the exact full-vertex can be represented in the simplified form in Eq.~(\ref{eq:2024-06-23-16-24}) when we can obtain the fully-irreducible vertex $\Lambda$ and the reducible vertex $\Phi_{l}$ in each channel $l$ exactly, which results in $\Psi_{l}=\Phi_{l}-V^{\text{non}-C}_{l}$ and $C=0$.
Hence, we can also construct approximations closer to the exact solution while considering the balance with the computational cost within the simplified form in Eq.~(\ref{eq:2024-06-23-16-24}). 

There exist other approaches that reduce the computational cost of calculating the two-particle quantities by decoupling the frequency variables~\cite{PhysRevB.99.165134,PhysRevB.100.205115}.
They are in different contexts from this work but similar to our simplification.
Especially in Ref.~\cite{PhysRevB.99.165134}, the authors show that the decoupling of the frequency variables becomes highly accurate when the interaction is strong.
This is consistent with our consideration in Sect.~\ref{sec:2024-07-11-13-13}.

\begin{table*}[t]
  {\includegraphics[width=170mm,clip]{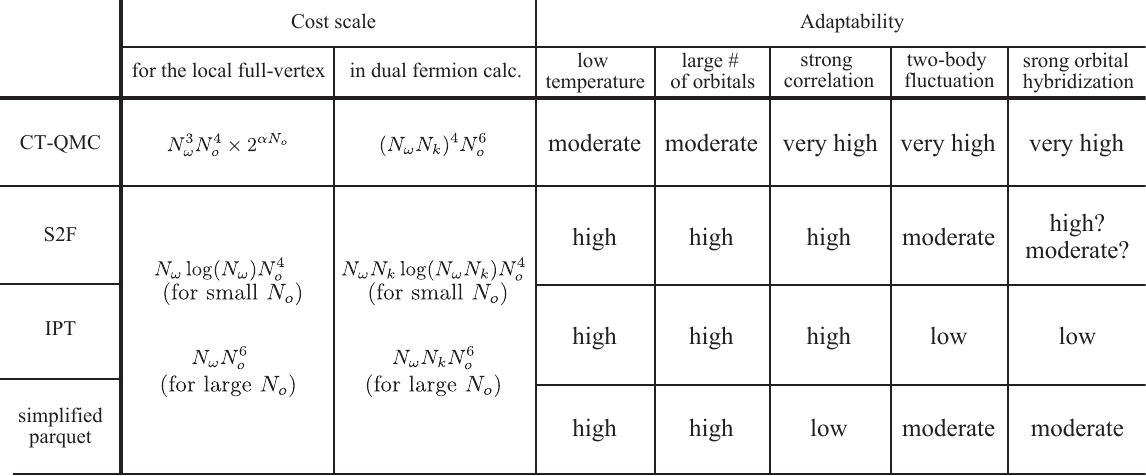}} 
  \caption{  
    Comparison of methods.
    $N_{\omega}, N_{k}, N_{o}$ are the number of frequencies, $k$-meshes, and orbitals, respectively.
    We assume that EDF is applied when combining S2F, IPT, and simplified parquet with the dual fermion calculation.
  }
  \label{tab:2025-02-23-16-59}
\end{table*}

\section{Conclusion}\label{sec:2024-07-11-13-15}

We have improved the simplification method for the local two-particle full-vertex we developed in the previous study~\cite{doi:10.7566/JPSJ.91.034002} by reconstructing the formulation so that we can treat the characteristic frequency structures of the full-vertex more correctly.
We have also improved the self-energy to the full-vertex (S2F) method for the improved full-vertex. 
We could have obtained the simplified local two-particle full-vertex qualitatively consistent with the numerically exact one while keeping the highly inexpensive numerical cost.
Our future work is developing numerically inexpensive diagrammatic extensions for DMFT by combining the improved simplified full-vertex.
We expect that the improved simplification method for the full-vertex can be useful for analyzing various strongly correlated systems.

\begin{acknowledgements}
  This study has been supported by Grant-in-Aid for Early-Career Scientists (Grant No.JP23K13061), JST FOREST Program (Grant No. JPMJFR212P) and Grant-in-Aid for Scientific Research B (Grant No. JP24K01333).

\end{acknowledgements}

\appendix 

\section{Detail calculation of the reducible vertex} \label{sec:2024-11-12-16-03}

Here, we show the detail of the transformation of the reducible vertex $\Phi_{l}$.
We start with the same equation as Eq.~(\ref{eq:2024-03-22-00-52}). 
\begin{align}
  \Phi_{l} 
  =&
  -\Gamma_{l}\chi_{l}\Gamma_{l}
  \nonumber \\
  =&
  -(\Lambda+\gamma_{l}) \chi_{l} (\Lambda + \gamma_{l})
  \nonumber \\
  =&
  -\Lambda \chi_{l} \Lambda
  \underbrace{
    -\gamma_{l} \chi_{l} \Lambda
  }_{\rm (I)}
  \underbrace{
    -\Lambda_{l} \chi_{l} \gamma_{l}
  }_{\rm (II)}
  \underbrace{
    -\gamma_{l} \chi_{l} \gamma_{l}
  }_{\rm (III)},
  \label{eq:2024-10-16-15-45}
\end{align}
where $\Gamma_{l}=\Lambda + \gamma_{l}$ and $\gamma_{l} = \Phi_{l_{1}} + \Phi_{l_{2}} \ (l\neq l_{1} \neq l_{2})$.
By using the relation $\chi_{l}=\chi_{0} - \chi_{0}\Gamma_{l}\chi_{l}$, we can transform the term~(I) as follows
\begin{align}
  {\rm (I)}
  =&
  -\gamma_{l}\chi_{0}\Lambda + \gamma_{l}\chi_{0}\Gamma_{l}\chi_{l}\Lambda
  \nonumber \\
  =&
  (-\gamma_{l}\chi_{0}\Gamma_{l}\Lambda^{-1})\underbrace{\Lambda \Gamma_{l}^{-1}}_{1-\gamma_{l}\Gamma_{l}^{-1}}\Lambda  - (-\gamma_{l}\chi_{0}\Gamma_{l}\Lambda^{-1})\Lambda \chi_{l}\Lambda
  \nonumber \\
  =&
  (-\gamma_{l}\chi_{0}\Gamma_{l}\Lambda^{-1})\Lambda - (-\gamma_{l}\chi_{0}\Gamma_{l}\Lambda)\Lambda\chi_{l}\Lambda 
  \nonumber \\
  &\hspace{60pt}
  - (-\gamma_{l}\chi_{0}\Gamma_{l}\Lambda^{-1})\gamma_{l}\Gamma_{l}^{-1}\Lambda
  \label{eq:2024-10-16-15-46}
\end{align}
Similarly, the term~(II) as
\begin{align}
  {\rm (II)}
  =&
  \Lambda(-\Lambda^{-1}\Gamma_{l}\chi_{0}\gamma_{l}) - \Lambda\chi_{l}\Lambda (-\Lambda^{-1}\Gamma_{l}\chi_{0}\gamma_{l})
  \nonumber \\
  &\hspace{60pt}
  - \Lambda\Gamma_{l}^{-1}\gamma_{l}(-\Lambda^{-1}\Gamma_{l}\chi_{0}\gamma_{l}),
  \label{eq:2024-10-16-15-47}
\end{align}
and the term (III) as 
\begin{align}
  {\rm (III)} 
  =& -\gamma_{l}\chi_{0}\gamma_{l} + \gamma_{l}\chi_{0}\Gamma_{l}\chi_{0}\gamma_{l} - \gamma_{l}\chi_{0}\Gamma_{l}\chi_{l}\Gamma_{l}\chi_{0}\gamma_{l}
  \nonumber \\
  =& -\gamma_{l}\chi_{0}\gamma_{l} + \gamma_{l}\chi_{0}\Gamma_{l}\chi_{0}\gamma_{l} \nonumber \\
  & - (-\gamma_{l}\chi_{0}\Gamma_{l}\Lambda^{-1})\Lambda\chi_{l}\Lambda(-\Lambda^{-1}\Gamma_{l}\chi_{0}\gamma_{l})
  \nonumber \\
  =& (-\gamma_{l}\chi_{0}\Gamma_{l}\Lambda^{-1}) \Lambda  (-\Lambda^{-1}\Gamma_{l}\chi_{0}\gamma_{l}) \nonumber \\
  &- (-\gamma_{l}\chi_{0}\Gamma_{l}\Lambda^{-1}) \Lambda\chi_{l}\Lambda  (-\Lambda^{-1}\Gamma_{l}\chi_{0}\gamma_{l}) 
  \nonumber \\
  & -\gamma_{l}\chi_{0}\gamma_{l} - (-\gamma_{l}\chi_{0}\Gamma_{l}\Lambda^{-1}) \Lambda\Gamma_{l}^{-1}\gamma_{l}  (-\Lambda^{-1}\Gamma_{l}\chi_{0}\gamma_{l}),
  \label{eq:2024-11-12-13-41}
\end{align}
where we use the following. 
\begin{align}
  \gamma_{l}\chi_{0}&\Gamma_{l}\chi_{0}\gamma_{l} \nonumber \\
  =& (-\gamma_{l}\chi_{0}\Gamma_{l}\Lambda^{-1}) \Lambda \underbrace{ \Gamma_{l}^{-1}\Lambda }_{1-\Gamma_{l}^{-1}\gamma_{l}} (-\Lambda^{-1}\Gamma_{l}\chi_{0}\gamma_{l}) \nonumber \\
  =& (-\gamma_{l}\chi_{0}\Gamma_{l}\Lambda^{-1}) \Lambda  (-\Lambda^{-1}\Gamma_{l}\chi_{0}\gamma_{l}) \nonumber \\ 
  &- (-\gamma_{l}\chi_{0}\Gamma_{l}\Lambda^{-1}) \Lambda\Gamma_{l}^{-1}\gamma_{l}  (-\Lambda^{-1}\Gamma_{l}\chi_{0}\gamma_{l}).  
  \label{eq:2024-12-18-13-09}
\end{align}
Then we obtain
\begin{align}
  \Phi_{l}
  =&
  (-\gamma_{l}\chi_{0}\Gamma_{l}\Lambda^{-1}) \Lambda 
  +  \Lambda  (-\Lambda^{-1}\Gamma_{l}\chi_{0}\gamma_{l}) \nonumber \\
  &+ (-\gamma_{l}\chi_{0}\Gamma_{l}\Lambda^{-1}) \Lambda  (-\Lambda^{-1}\Gamma_{l}\chi_{0}\gamma_{l}) \nonumber \\
  &+ (-\gamma_{l}\chi_{0}\Gamma_{l}\Lambda^{-1}) \Psi_{l}
  +  \Psi_{l}  (-\Lambda^{-1}\Gamma_{l}\chi_{0}\gamma_{l}) \nonumber \\
  &+ (-\gamma_{l}\chi_{0}\Gamma_{l}\Lambda^{-1}) \Psi_{l}  (-\Lambda^{-1}\Gamma_{l}\chi_{0}\gamma_{l})
  \nonumber \\
  \nonumber \\
  & 
  \begin{rcases}
    +& \gamma_{l}\chi_{0}\Gamma_{l}\Lambda\gamma_{l}\Gamma_{l}^{-1}\Lambda \\ 
    +& \Lambda\Gamma_{l}^{-1}\gamma_{l}\Lambda\Gamma_{l}\chi_{0}\gamma_{l} \\
    -&\gamma_{l}\chi_{0}\gamma_{l} - \gamma_{l}\chi_{0}\gamma_{l}\Lambda^{-1}\Gamma_{l}\chi_{0}\gamma_{l} 
  \end{rcases}
  {\rm (IV)}
  \label{eq:2024-11-12-15-54}
\end{align}
where $\Psi_{l}=-\Lambda\chi_{l}\Lambda$.
We can transform the term (IV) by using $\gamma_{l}=\Gamma_{l}-\Lambda$ as
\begin{align}
  {\rm (IV)} 
  =&
  (\Gamma_{l}-\Lambda) \chi_{0}\Gamma_{l}\Lambda^{-1}(\Gamma_{l}-\Lambda)\Gamma_{l}^{-1}\Lambda \nonumber \\
  &+ \Lambda\Gamma_{l}^{-1}(\Gamma_{l}-\Lambda)\Lambda\Gamma_{l}\chi_{0}(\Gamma_{l}-\Lambda)
  \nonumber \\
  &- (\Gamma_{l}-\Lambda) \chi_{0} (\Gamma_{l}-\Lambda) \nonumber \\
  &- (\Gamma_{l}-\Lambda) \chi_{0} (\Gamma_{l}-\Lambda) \Lambda^{-1}\Gamma_{l}\chi_{0}(\Gamma_{l}-\Lambda)
  \nonumber \\
  =&
  (\Gamma_{l}-\Lambda) \chi_{0} (\Gamma_{l}-\Lambda) \nonumber \\
  &+ (\Gamma_{l}-\Lambda) \chi_{0} (\Gamma_{l}-\Lambda)
  \nonumber \\
  &- (\Gamma_{l}-\Lambda) \chi_{0} (\Gamma_{l}-\Lambda) \nonumber \\
  &- (\Gamma_{l}-\Lambda) \chi_{0} (\Gamma_{l}-\Lambda) \Lambda^{-1}\Gamma_{l}\chi_{0}(\Gamma_{l}-\Lambda)
  \nonumber \\
  =&
  (\Gamma_{l}-\Lambda) [ \chi_{0} -  \chi_{0}(\Gamma_{l}\Lambda^{-1}-1) \Gamma_{l}\chi_{0} ] (\Gamma_{l}-\Lambda).
  \label{eq:2023-09-26-01-54}
\end{align}
By substituting Eq.~(\ref{eq:2023-09-26-01-54}) to Eq.~(\ref{eq:2024-11-12-15-54}), we can obtain Eq.~(\ref{eq:2024-05-19-14-46}).

\bibliographystyle{apsrev4-2_edited}


\begin{thebibliography}{52}%
\makeatletter
\providecommand \@ifxundefined [1]{%
 \@ifx{#1\undefined}
}%
\providecommand \@ifnum [1]{%
 \ifnum #1\expandafter \@firstoftwo
 \else \expandafter \@secondoftwo
 \fi
}%
\providecommand \@ifx [1]{%
 \ifx #1\expandafter \@firstoftwo
 \else \expandafter \@secondoftwo
 \fi
}%
\providecommand \natexlab [1]{#1}%
\providecommand \enquote  [1]{``#1''}%
\providecommand \bibnamefont  [1]{#1}%
\providecommand \bibfnamefont [1]{#1}%
\providecommand \citenamefont [1]{#1}%
\providecommand \href@noop [0]{\@secondoftwo}%
\providecommand \href [0]{\begingroup \@sanitize@url \@href}%
\providecommand \@href[1]{\@@startlink{#1}\@@href}%
\providecommand \@@href[1]{\endgroup#1\@@endlink}%
\providecommand \@sanitize@url [0]{\catcode `\\12\catcode `\$12\catcode
  `\&12\catcode `\#12\catcode `\^12\catcode `\_12\catcode `\%12\relax}%
\providecommand \@@startlink[1]{}%
\providecommand \@@endlink[0]{}%
\providecommand \url  [0]{\begingroup\@sanitize@url \@url }%
\providecommand \@url [1]{\endgroup\@href {#1}{\urlprefix }}%
\providecommand \urlprefix  [0]{URL }%
\providecommand \Eprint [0]{\href }%
\providecommand \doibase [0]{https://doi.org/}%
\providecommand \selectlanguage [0]{\@gobble}%
\providecommand \bibinfo  [0]{\@secondoftwo}%
\providecommand \bibfield  [0]{\@secondoftwo}%
\providecommand \translation [1]{[#1]}%
\providecommand \BibitemOpen [0]{}%
\providecommand \bibitemStop [0]{}%
\providecommand \bibitemNoStop [0]{.\EOS\space}%
\providecommand \EOS [0]{\spacefactor3000\relax}%
\providecommand \BibitemShut  [1]{\csname bibitem#1\endcsname}%
\let\auto@bib@innerbib\@empty
\bibitem [{\citenamefont {Mizuno}\ \emph {et~al.}(2022)\citenamefont {Mizuno},
  \citenamefont {Ochi},\ and\ \citenamefont
  {Kuroki}}]{doi:10.7566/JPSJ.91.034002}%
  \BibitemOpen
  \bibfield  {author} {\bibinfo {author} {\bibfnamefont {R.}~\bibnamefont
  {Mizuno}}, \bibinfo {author} {\bibfnamefont {M.}~\bibnamefont {Ochi}},\ and\
  \bibinfo {author} {\bibfnamefont {K.}~\bibnamefont {Kuroki}},\ }\bibinfo
  {title} {Simplification of the local full vertex in the impurity problem in
  dmft and its applications for the nonlocal correlation},\ \href
  {https://doi.org/10.7566/JPSJ.91.034002} {\bibfield  {journal} {\bibinfo
  {journal} {Journal of the Physical Society of Japan}\ }\textbf {\bibinfo
  {volume} {91}},\ \bibinfo {pages} {034002} (\bibinfo {year} {2022})},\
  \Eprint {https://arxiv.org/abs/https://doi.org/10.7566/JPSJ.91.034002}
  {https://doi.org/10.7566/JPSJ.91.034002} \BibitemShut {NoStop}%
\bibitem [{\citenamefont {Georges}\ \emph {et~al.}(1996)\citenamefont
  {Georges}, \citenamefont {Kotliar}, \citenamefont {Krauth},\ and\
  \citenamefont {Rozenberg}}]{RevModPhys.68.13}%
  \BibitemOpen
  \bibfield  {author} {\bibinfo {author} {\bibfnamefont {A.}~\bibnamefont
  {Georges}}, \bibinfo {author} {\bibfnamefont {G.}~\bibnamefont {Kotliar}},
  \bibinfo {author} {\bibfnamefont {W.}~\bibnamefont {Krauth}},\ and\ \bibinfo
  {author} {\bibfnamefont {M.~J.}\ \bibnamefont {Rozenberg}},\ }\bibinfo
  {title} {Dynamical mean-field theory of strongly correlated fermion systems
  and the limit of infinite dimensions},\ \href
  {https://doi.org/10.1103/RevModPhys.68.13} {\bibfield  {journal} {\bibinfo
  {journal} {Rev. Mod. Phys.}\ }\textbf {\bibinfo {volume} {68}},\ \bibinfo
  {pages} {13} (\bibinfo {year} {1996})}\BibitemShut {NoStop}%
\bibitem [{\citenamefont {Maier}\ \emph {et~al.}(2005)\citenamefont {Maier},
  \citenamefont {Jarrell}, \citenamefont {Pruschke},\ and\ \citenamefont
  {Hettler}}]{RevModPhys.77.1027}%
  \BibitemOpen
  \bibfield  {author} {\bibinfo {author} {\bibfnamefont {T.}~\bibnamefont
  {Maier}}, \bibinfo {author} {\bibfnamefont {M.}~\bibnamefont {Jarrell}},
  \bibinfo {author} {\bibfnamefont {T.}~\bibnamefont {Pruschke}},\ and\
  \bibinfo {author} {\bibfnamefont {M.~H.}\ \bibnamefont {Hettler}},\ }\bibinfo
  {title} {Quantum cluster theories},\ \href
  {https://doi.org/10.1103/RevModPhys.77.1027} {\bibfield  {journal} {\bibinfo
  {journal} {Rev. Mod. Phys.}\ }\textbf {\bibinfo {volume} {77}},\ \bibinfo
  {pages} {1027} (\bibinfo {year} {2005})}\BibitemShut {NoStop}%
\bibitem [{\citenamefont {Rohringer}\ \emph {et~al.}(2018)\citenamefont
  {Rohringer}, \citenamefont {Hafermann}, \citenamefont {Toschi}, \citenamefont
  {Katanin}, \citenamefont {Antipov}, \citenamefont {Katsnelson}, \citenamefont
  {Lichtenstein}, \citenamefont {Rubtsov},\ and\ \citenamefont
  {Held}}]{RevModPhys.90.025003}%
  \BibitemOpen
  \bibfield  {author} {\bibinfo {author} {\bibfnamefont {G.}~\bibnamefont
  {Rohringer}}, \bibinfo {author} {\bibfnamefont {H.}~\bibnamefont
  {Hafermann}}, \bibinfo {author} {\bibfnamefont {A.}~\bibnamefont {Toschi}},
  \bibinfo {author} {\bibfnamefont {A.~A.}\ \bibnamefont {Katanin}}, \bibinfo
  {author} {\bibfnamefont {A.~E.}\ \bibnamefont {Antipov}}, \bibinfo {author}
  {\bibfnamefont {M.~I.}\ \bibnamefont {Katsnelson}}, \bibinfo {author}
  {\bibfnamefont {A.~I.}\ \bibnamefont {Lichtenstein}}, \bibinfo {author}
  {\bibfnamefont {A.~N.}\ \bibnamefont {Rubtsov}},\ and\ \bibinfo {author}
  {\bibfnamefont {K.}~\bibnamefont {Held}},\ }\bibinfo {title} {Diagrammatic
  routes to nonlocal correlations beyond dynamical mean field theory},\ \href
  {https://doi.org/10.1103/RevModPhys.90.025003} {\bibfield  {journal}
  {\bibinfo  {journal} {Rev. Mod. Phys.}\ }\textbf {\bibinfo {volume} {90}},\
  \bibinfo {pages} {025003} (\bibinfo {year} {2018})}\BibitemShut {NoStop}%
\bibitem [{\citenamefont {Rubtsov}\ \emph {et~al.}(2008)\citenamefont
  {Rubtsov}, \citenamefont {Katsnelson},\ and\ \citenamefont
  {Lichtenstein}}]{PhysRevB.77.033101}%
  \BibitemOpen
  \bibfield  {author} {\bibinfo {author} {\bibfnamefont {A.~N.}\ \bibnamefont
  {Rubtsov}}, \bibinfo {author} {\bibfnamefont {M.~I.}\ \bibnamefont
  {Katsnelson}},\ and\ \bibinfo {author} {\bibfnamefont {A.~I.}\ \bibnamefont
  {Lichtenstein}},\ }\bibinfo {title} {Dual fermion approach to nonlocal
  correlations in the hubbard model},\ \href
  {https://doi.org/10.1103/PhysRevB.77.033101} {\bibfield  {journal} {\bibinfo
  {journal} {Phys. Rev. B}\ }\textbf {\bibinfo {volume} {77}},\ \bibinfo
  {pages} {033101} (\bibinfo {year} {2008})}\BibitemShut {NoStop}%
\bibitem [{\citenamefont {Rubtsov}\ \emph {et~al.}(2009)\citenamefont
  {Rubtsov}, \citenamefont {Katsnelson}, \citenamefont {Lichtenstein},\ and\
  \citenamefont {Georges}}]{PhysRevB.79.045133}%
  \BibitemOpen
  \bibfield  {author} {\bibinfo {author} {\bibfnamefont {A.~N.}\ \bibnamefont
  {Rubtsov}}, \bibinfo {author} {\bibfnamefont {M.~I.}\ \bibnamefont
  {Katsnelson}}, \bibinfo {author} {\bibfnamefont {A.~I.}\ \bibnamefont
  {Lichtenstein}},\ and\ \bibinfo {author} {\bibfnamefont {A.}~\bibnamefont
  {Georges}},\ }\bibinfo {title} {Dual fermion approach to the two-dimensional
  hubbard model: Antiferromagnetic fluctuations and fermi arcs},\ \href
  {https://doi.org/10.1103/PhysRevB.79.045133} {\bibfield  {journal} {\bibinfo
  {journal} {Phys. Rev. B}\ }\textbf {\bibinfo {volume} {79}},\ \bibinfo
  {pages} {045133} (\bibinfo {year} {2009})}\BibitemShut {NoStop}%
\bibitem [{\citenamefont {Otsuki}\ \emph {et~al.}(2014)\citenamefont {Otsuki},
  \citenamefont {Hafermann},\ and\ \citenamefont
  {Lichtenstein}}]{PhysRevB.90.235132}%
  \BibitemOpen
  \bibfield  {author} {\bibinfo {author} {\bibfnamefont {J.}~\bibnamefont
  {Otsuki}}, \bibinfo {author} {\bibfnamefont {H.}~\bibnamefont {Hafermann}},\
  and\ \bibinfo {author} {\bibfnamefont {A.~I.}\ \bibnamefont {Lichtenstein}},\
  }\bibinfo {title} {Superconductivity, antiferromagnetism, and phase
  separation in the two-dimensional hubbard model: A dual-fermion approach},\
  \href {https://doi.org/10.1103/PhysRevB.90.235132} {\bibfield  {journal}
  {\bibinfo  {journal} {Phys. Rev. B}\ }\textbf {\bibinfo {volume} {90}},\
  \bibinfo {pages} {235132} (\bibinfo {year} {2014})}\BibitemShut {NoStop}%
\bibitem [{\citenamefont {Hirschmeier}\ \emph {et~al.}(2018)\citenamefont
  {Hirschmeier}, \citenamefont {Hafermann},\ and\ \citenamefont
  {Lichtenstein}}]{PhysRevB.97.115150}%
  \BibitemOpen
  \bibfield  {author} {\bibinfo {author} {\bibfnamefont {D.}~\bibnamefont
  {Hirschmeier}}, \bibinfo {author} {\bibfnamefont {H.}~\bibnamefont
  {Hafermann}},\ and\ \bibinfo {author} {\bibfnamefont {A.~I.}\ \bibnamefont
  {Lichtenstein}},\ }\bibinfo {title} {Multiband dual fermion approach to
  quantum criticality in the hubbard honeycomb lattice},\ \href
  {https://doi.org/10.1103/PhysRevB.97.115150} {\bibfield  {journal} {\bibinfo
  {journal} {Phys. Rev. B}\ }\textbf {\bibinfo {volume} {97}},\ \bibinfo
  {pages} {115150} (\bibinfo {year} {2018})}\BibitemShut {NoStop}%
\bibitem [{\citenamefont {van Loon}\ \emph {et~al.}(2018)\citenamefont {van
  Loon}, \citenamefont {Katsnelson},\ and\ \citenamefont
  {Hafermann}}]{PhysRevB.98.155117}%
  \BibitemOpen
  \bibfield  {author} {\bibinfo {author} {\bibfnamefont {E.~G. C.~P.}\
  \bibnamefont {van Loon}}, \bibinfo {author} {\bibfnamefont {M.~I.}\
  \bibnamefont {Katsnelson}},\ and\ \bibinfo {author} {\bibfnamefont
  {H.}~\bibnamefont {Hafermann}},\ }\bibinfo {title} {Second-order dual fermion
  approach to the mott transition in the two-dimensional hubbard model},\ \href
  {https://doi.org/10.1103/PhysRevB.98.155117} {\bibfield  {journal} {\bibinfo
  {journal} {Phys. Rev. B}\ }\textbf {\bibinfo {volume} {98}},\ \bibinfo
  {pages} {155117} (\bibinfo {year} {2018})}\BibitemShut {NoStop}%
\bibitem [{\citenamefont {Toschi}\ \emph {et~al.}(2007)\citenamefont {Toschi},
  \citenamefont {Katanin},\ and\ \citenamefont {Held}}]{PhysRevB.75.045118}%
  \BibitemOpen
  \bibfield  {author} {\bibinfo {author} {\bibfnamefont {A.}~\bibnamefont
  {Toschi}}, \bibinfo {author} {\bibfnamefont {A.~A.}\ \bibnamefont
  {Katanin}},\ and\ \bibinfo {author} {\bibfnamefont {K.}~\bibnamefont
  {Held}},\ }\bibinfo {title} {Dynamical vertex approximation: A step beyond
  dynamical mean-field theory},\ \href
  {https://doi.org/10.1103/PhysRevB.75.045118} {\bibfield  {journal} {\bibinfo
  {journal} {Phys. Rev. B}\ }\textbf {\bibinfo {volume} {75}},\ \bibinfo
  {pages} {045118} (\bibinfo {year} {2007})}\BibitemShut {NoStop}%
\bibitem [{\citenamefont {Toschi}\ \emph {et~al.}(2011)\citenamefont {Toschi},
  \citenamefont {Rohringer}, \citenamefont {Katanin},\ and\ \citenamefont
  {Held}}]{andp.201100036}%
  \BibitemOpen
  \bibfield  {author} {\bibinfo {author} {\bibfnamefont {A.}~\bibnamefont
  {Toschi}}, \bibinfo {author} {\bibfnamefont {G.}~\bibnamefont {Rohringer}},
  \bibinfo {author} {\bibfnamefont {A.}~\bibnamefont {Katanin}},\ and\ \bibinfo
  {author} {\bibfnamefont {K.}~\bibnamefont {Held}},\ }\bibinfo {title} {Ab
  initio calculations with the dynamical vertex approximation},\ \href
  {https://doi.org/https://doi.org/10.1002/andp.201100036} {\bibfield
  {journal} {\bibinfo  {journal} {Annalen der Physik}\ }\textbf {\bibinfo
  {volume} {523}},\ \bibinfo {pages} {698} (\bibinfo {year} {2011})},\ \Eprint
  {https://arxiv.org/abs/https://onlinelibrary.wiley.com/doi/pdf/10.1002/andp.201100036}
  {https://onlinelibrary.wiley.com/doi/pdf/10.1002/andp.201100036} \BibitemShut
  {NoStop}%
\bibitem [{\citenamefont {Kusunose}(2006)}]{doi:10.1143/JPSJ.75.054713}%
  \BibitemOpen
  \bibfield  {author} {\bibinfo {author} {\bibfnamefont {H.}~\bibnamefont
  {Kusunose}},\ }\bibinfo {title} {Influence of spatial correlations in
  strongly correlated electron systems: Extension to dynamical mean field
  approximation},\ \href {https://doi.org/10.1143/JPSJ.75.054713} {\bibfield
  {journal} {\bibinfo  {journal} {Journal of the Physical Society of Japan}\
  }\textbf {\bibinfo {volume} {75}},\ \bibinfo {pages} {054713} (\bibinfo
  {year} {2006})},\ \Eprint
  {https://arxiv.org/abs/http://dx.doi.org/10.1143/JPSJ.75.054713}
  {http://dx.doi.org/10.1143/JPSJ.75.054713} \BibitemShut {NoStop}%
\bibitem [{\citenamefont {Rubtsov}\ \emph {et~al.}(2005)\citenamefont
  {Rubtsov}, \citenamefont {Savkin},\ and\ \citenamefont
  {Lichtenstein}}]{PhysRevB.72.035122}%
  \BibitemOpen
  \bibfield  {author} {\bibinfo {author} {\bibfnamefont {A.~N.}\ \bibnamefont
  {Rubtsov}}, \bibinfo {author} {\bibfnamefont {V.~V.}\ \bibnamefont
  {Savkin}},\ and\ \bibinfo {author} {\bibfnamefont {A.~I.}\ \bibnamefont
  {Lichtenstein}},\ }\bibinfo {title} {Continuous-time quantum monte carlo
  method for fermions},\ \href {https://doi.org/10.1103/PhysRevB.72.035122}
  {\bibfield  {journal} {\bibinfo  {journal} {Phys. Rev. B}\ }\textbf {\bibinfo
  {volume} {72}},\ \bibinfo {pages} {035122} (\bibinfo {year}
  {2005})}\BibitemShut {NoStop}%
\bibitem [{\citenamefont {Werner}\ \emph {et~al.}(2006)\citenamefont {Werner},
  \citenamefont {Comanac}, \citenamefont {de' Medici}, \citenamefont {Troyer},\
  and\ \citenamefont {Millis}}]{PhysRevLett.97.076405}%
  \BibitemOpen
  \bibfield  {author} {\bibinfo {author} {\bibfnamefont {P.}~\bibnamefont
  {Werner}}, \bibinfo {author} {\bibfnamefont {A.}~\bibnamefont {Comanac}},
  \bibinfo {author} {\bibfnamefont {L.}~\bibnamefont {de' Medici}}, \bibinfo
  {author} {\bibfnamefont {M.}~\bibnamefont {Troyer}},\ and\ \bibinfo {author}
  {\bibfnamefont {A.~J.}\ \bibnamefont {Millis}},\ }\bibinfo {title}
  {Continuous-time solver for quantum impurity models},\ \href
  {https://doi.org/10.1103/PhysRevLett.97.076405} {\bibfield  {journal}
  {\bibinfo  {journal} {Phys. Rev. Lett.}\ }\textbf {\bibinfo {volume} {97}},\
  \bibinfo {pages} {076405} (\bibinfo {year} {2006})}\BibitemShut {NoStop}%
\bibitem [{\citenamefont {Gull}\ \emph {et~al.}(2007)\citenamefont {Gull},
  \citenamefont {Werner}, \citenamefont {Millis},\ and\ \citenamefont
  {Troyer}}]{PhysRevB.76.235123}%
  \BibitemOpen
  \bibfield  {author} {\bibinfo {author} {\bibfnamefont {E.}~\bibnamefont
  {Gull}}, \bibinfo {author} {\bibfnamefont {P.}~\bibnamefont {Werner}},
  \bibinfo {author} {\bibfnamefont {A.}~\bibnamefont {Millis}},\ and\ \bibinfo
  {author} {\bibfnamefont {M.}~\bibnamefont {Troyer}},\ }\bibinfo {title}
  {Performance analysis of continuous-time solvers for quantum impurity
  models},\ \href {https://doi.org/10.1103/PhysRevB.76.235123} {\bibfield
  {journal} {\bibinfo  {journal} {Phys. Rev. B}\ }\textbf {\bibinfo {volume}
  {76}},\ \bibinfo {pages} {235123} (\bibinfo {year} {2007})}\BibitemShut
  {NoStop}%
\bibitem [{\citenamefont {Werner}\ and\ \citenamefont
  {Millis}(2006)}]{PhysRevB.74.155107}%
  \BibitemOpen
  \bibfield  {author} {\bibinfo {author} {\bibfnamefont {P.}~\bibnamefont
  {Werner}}\ and\ \bibinfo {author} {\bibfnamefont {A.~J.}\ \bibnamefont
  {Millis}},\ }\bibinfo {title} {Hybridization expansion impurity solver:
  General formulation and application to kondo lattice and two-orbital
  models},\ \href {https://doi.org/10.1103/PhysRevB.74.155107} {\bibfield
  {journal} {\bibinfo  {journal} {Phys. Rev. B}\ }\textbf {\bibinfo {volume}
  {74}},\ \bibinfo {pages} {155107} (\bibinfo {year} {2006})}\BibitemShut
  {NoStop}%
\bibitem [{\citenamefont {Otsuki}\ \emph {et~al.}(2007)\citenamefont {Otsuki},
  \citenamefont {Kusunose}, \citenamefont {Werner},\ and\ \citenamefont
  {Kuramoto}}]{doi:10.1143/JPSJ.76.114707}%
  \BibitemOpen
  \bibfield  {author} {\bibinfo {author} {\bibfnamefont {J.}~\bibnamefont
  {Otsuki}}, \bibinfo {author} {\bibfnamefont {H.}~\bibnamefont {Kusunose}},
  \bibinfo {author} {\bibfnamefont {P.}~\bibnamefont {Werner}},\ and\ \bibinfo
  {author} {\bibfnamefont {Y.}~\bibnamefont {Kuramoto}},\ }\bibinfo {title}
  {Continuous-time quantum monte carlo method for the coqblin-schrieffer
  model},\ \href {https://doi.org/10.1143/JPSJ.76.114707} {\bibfield  {journal}
  {\bibinfo  {journal} {Journal of the Physical Society of Japan}\ }\textbf
  {\bibinfo {volume} {76}},\ \bibinfo {pages} {114707} (\bibinfo {year}
  {2007})},\ \Eprint
  {https://arxiv.org/abs/https://doi.org/10.1143/JPSJ.76.114707}
  {https://doi.org/10.1143/JPSJ.76.114707} \BibitemShut {NoStop}%
\bibitem [{\citenamefont {Caffarel}\ and\ \citenamefont
  {Krauth}(1994)}]{PhysRevLett.72.1545}%
  \BibitemOpen
  \bibfield  {author} {\bibinfo {author} {\bibfnamefont {M.}~\bibnamefont
  {Caffarel}}\ and\ \bibinfo {author} {\bibfnamefont {W.}~\bibnamefont
  {Krauth}},\ }\bibinfo {title} {Exact diagonalization approach to correlated
  fermions in infinite dimensions: Mott transition and superconductivity},\
  \href {https://doi.org/10.1103/PhysRevLett.72.1545} {\bibfield  {journal}
  {\bibinfo  {journal} {Phys. Rev. Lett.}\ }\textbf {\bibinfo {volume} {72}},\
  \bibinfo {pages} {1545} (\bibinfo {year} {1994})}\BibitemShut {NoStop}%
\bibitem [{\citenamefont {Zgid}\ \emph {et~al.}(2012)\citenamefont {Zgid},
  \citenamefont {Gull},\ and\ \citenamefont {Chan}}]{PhysRevB.86.165128}%
  \BibitemOpen
  \bibfield  {author} {\bibinfo {author} {\bibfnamefont {D.}~\bibnamefont
  {Zgid}}, \bibinfo {author} {\bibfnamefont {E.}~\bibnamefont {Gull}},\ and\
  \bibinfo {author} {\bibfnamefont {G.~K.-L.}\ \bibnamefont {Chan}},\ }\bibinfo
  {title} {Truncated configuration interaction expansions as solvers for
  correlated quantum impurity models and dynamical mean-field theory},\ \href
  {https://doi.org/10.1103/PhysRevB.86.165128} {\bibfield  {journal} {\bibinfo
  {journal} {Phys. Rev. B}\ }\textbf {\bibinfo {volume} {86}},\ \bibinfo
  {pages} {165128} (\bibinfo {year} {2012})}\BibitemShut {NoStop}%
\bibitem [{\citenamefont {Kugler}\ \emph {et~al.}(2021)\citenamefont {Kugler},
  \citenamefont {Lee},\ and\ \citenamefont {von Delft}}]{PhysRevX.11.041006}%
  \BibitemOpen
  \bibfield  {author} {\bibinfo {author} {\bibfnamefont {F.~B.}\ \bibnamefont
  {Kugler}}, \bibinfo {author} {\bibfnamefont {S.-S.~B.}\ \bibnamefont {Lee}},\
  and\ \bibinfo {author} {\bibfnamefont {J.}~\bibnamefont {von Delft}},\
  }\bibinfo {title} {Multipoint correlation functions: Spectral representation
  and numerical evaluation},\ \href
  {https://doi.org/10.1103/PhysRevX.11.041006} {\bibfield  {journal} {\bibinfo
  {journal} {Phys. Rev. X}\ }\textbf {\bibinfo {volume} {11}},\ \bibinfo
  {pages} {041006} (\bibinfo {year} {2021})}\BibitemShut {NoStop}%
\bibitem [{\citenamefont {Gunacker}\ \emph {et~al.}(2016)\citenamefont
  {Gunacker}, \citenamefont {Wallerberger}, \citenamefont {Ribic},
  \citenamefont {Hausoel}, \citenamefont {Sangiovanni},\ and\ \citenamefont
  {Held}}]{PhysRevB.94.125153}%
  \BibitemOpen
  \bibfield  {author} {\bibinfo {author} {\bibfnamefont {P.}~\bibnamefont
  {Gunacker}}, \bibinfo {author} {\bibfnamefont {M.}~\bibnamefont
  {Wallerberger}}, \bibinfo {author} {\bibfnamefont {T.}~\bibnamefont {Ribic}},
  \bibinfo {author} {\bibfnamefont {A.}~\bibnamefont {Hausoel}}, \bibinfo
  {author} {\bibfnamefont {G.}~\bibnamefont {Sangiovanni}},\ and\ \bibinfo
  {author} {\bibfnamefont {K.}~\bibnamefont {Held}},\ }\bibinfo {title}
  {Worm-improved estimators in continuous-time quantum monte carlo},\ \href
  {https://doi.org/10.1103/PhysRevB.94.125153} {\bibfield  {journal} {\bibinfo
  {journal} {Phys. Rev. B}\ }\textbf {\bibinfo {volume} {94}},\ \bibinfo
  {pages} {125153} (\bibinfo {year} {2016})}\BibitemShut {NoStop}%
\bibitem [{\citenamefont {Shinaoka}\ \emph {et~al.}(2020)\citenamefont
  {Shinaoka}, \citenamefont {Geffroy}, \citenamefont {Wallerberger},
  \citenamefont {Otsuki}, \citenamefont {Yoshimi}, \citenamefont {Gull},\ and\
  \citenamefont {Kunes}}]{10.21468/SciPostPhys.8.1.012}%
  \BibitemOpen
  \bibfield  {author} {\bibinfo {author} {\bibfnamefont {H.}~\bibnamefont
  {Shinaoka}}, \bibinfo {author} {\bibfnamefont {D.}~\bibnamefont {Geffroy}},
  \bibinfo {author} {\bibfnamefont {M.}~\bibnamefont {Wallerberger}}, \bibinfo
  {author} {\bibfnamefont {J.}~\bibnamefont {Otsuki}}, \bibinfo {author}
  {\bibfnamefont {K.}~\bibnamefont {Yoshimi}}, \bibinfo {author} {\bibfnamefont
  {E.}~\bibnamefont {Gull}},\ and\ \bibinfo {author} {\bibfnamefont
  {J.}~\bibnamefont {Kunes}},\ }\bibinfo {title} {{Sparse sampling and tensor
  network representation of two-particle Green's functions}},\ \href
  {https://doi.org/10.21468/SciPostPhys.8.1.012} {\bibfield  {journal}
  {\bibinfo  {journal} {SciPost Phys.}\ }\textbf {\bibinfo {volume} {8}},\
  \bibinfo {pages} {12} (\bibinfo {year} {2020})}\BibitemShut {NoStop}%
\bibitem [{\citenamefont {Moghadas}\ \emph {et~al.}(2024)\citenamefont
  {Moghadas}, \citenamefont {Dr{\"a}ger}, \citenamefont {Toschi}, \citenamefont
  {Zang}, \citenamefont {Medvidovi{\'{c}}}, \citenamefont {Kiese},
  \citenamefont {Millis}, \citenamefont {Sengupta}, \citenamefont
  {Andergassen},\ and\ \citenamefont {Di~Sante}}]{Moghadas2024}%
  \BibitemOpen
  \bibfield  {author} {\bibinfo {author} {\bibfnamefont {E.}~\bibnamefont
  {Moghadas}}, \bibinfo {author} {\bibfnamefont {N.}~\bibnamefont
  {Dr{\"a}ger}}, \bibinfo {author} {\bibfnamefont {A.}~\bibnamefont {Toschi}},
  \bibinfo {author} {\bibfnamefont {J.}~\bibnamefont {Zang}}, \bibinfo {author}
  {\bibfnamefont {M.}~\bibnamefont {Medvidovi{\'{c}}}}, \bibinfo {author}
  {\bibfnamefont {D.}~\bibnamefont {Kiese}}, \bibinfo {author} {\bibfnamefont
  {A.~J.}\ \bibnamefont {Millis}}, \bibinfo {author} {\bibfnamefont {A.~M.}\
  \bibnamefont {Sengupta}}, \bibinfo {author} {\bibfnamefont {S.}~\bibnamefont
  {Andergassen}},\ and\ \bibinfo {author} {\bibfnamefont {D.}~\bibnamefont
  {Di~Sante}},\ }\bibinfo {title} {Compressing the two-particle green's
  function using wavelets: Theory and application to the hubbard atom},\ \href
  {https://doi.org/10.1140/epjp/s13360-024-05403-9} {\bibfield  {journal}
  {\bibinfo  {journal} {The European Physical Journal Plus}\ }\textbf {\bibinfo
  {volume} {139}},\ \bibinfo {pages} {700} (\bibinfo {year}
  {2024})}\BibitemShut {NoStop}%
\bibitem [{\citenamefont {Ge}\ \emph {et~al.}(2024)\citenamefont {Ge},
  \citenamefont {Ritz}, \citenamefont {Walter}, \citenamefont {Aguirre},
  \citenamefont {von Delft},\ and\ \citenamefont
  {Kugler}}]{PhysRevB.109.115128}%
  \BibitemOpen
  \bibfield  {author} {\bibinfo {author} {\bibfnamefont {A.}~\bibnamefont
  {Ge}}, \bibinfo {author} {\bibfnamefont {N.}~\bibnamefont {Ritz}}, \bibinfo
  {author} {\bibfnamefont {E.}~\bibnamefont {Walter}}, \bibinfo {author}
  {\bibfnamefont {S.}~\bibnamefont {Aguirre}}, \bibinfo {author} {\bibfnamefont
  {J.}~\bibnamefont {von Delft}},\ and\ \bibinfo {author} {\bibfnamefont
  {F.~B.}\ \bibnamefont {Kugler}},\ }\bibinfo {title} {Real-frequency quantum
  field theory applied to the single-impurity anderson model},\ \href
  {https://doi.org/10.1103/PhysRevB.109.115128} {\bibfield  {journal} {\bibinfo
   {journal} {Phys. Rev. B}\ }\textbf {\bibinfo {volume} {109}},\ \bibinfo
  {pages} {115128} (\bibinfo {year} {2024})}\BibitemShut {NoStop}%
\bibitem [{\citenamefont {Kaufmann}\ \emph {et~al.}(2017)\citenamefont
  {Kaufmann}, \citenamefont {Gunacker},\ and\ \citenamefont
  {Held}}]{PhysRevB.96.035114}%
  \BibitemOpen
  \bibfield  {author} {\bibinfo {author} {\bibfnamefont {J.}~\bibnamefont
  {Kaufmann}}, \bibinfo {author} {\bibfnamefont {P.}~\bibnamefont {Gunacker}},\
  and\ \bibinfo {author} {\bibfnamefont {K.}~\bibnamefont {Held}},\ }\bibinfo
  {title} {Continuous-time quantum monte carlo calculation of multiorbital
  vertex asymptotics},\ \href {https://doi.org/10.1103/PhysRevB.96.035114}
  {\bibfield  {journal} {\bibinfo  {journal} {Phys. Rev. B}\ }\textbf {\bibinfo
  {volume} {96}},\ \bibinfo {pages} {035114} (\bibinfo {year}
  {2017})}\BibitemShut {NoStop}%
\bibitem [{\citenamefont {Rohringer}\ \emph {et~al.}(2012)\citenamefont
  {Rohringer}, \citenamefont {Valli},\ and\ \citenamefont
  {Toschi}}]{PhysRevB.86.125114}%
  \BibitemOpen
  \bibfield  {author} {\bibinfo {author} {\bibfnamefont {G.}~\bibnamefont
  {Rohringer}}, \bibinfo {author} {\bibfnamefont {A.}~\bibnamefont {Valli}},\
  and\ \bibinfo {author} {\bibfnamefont {A.}~\bibnamefont {Toschi}},\ }\bibinfo
  {title} {Local electronic correlation at the two-particle level},\ \href
  {https://doi.org/10.1103/PhysRevB.86.125114} {\bibfield  {journal} {\bibinfo
  {journal} {Phys. Rev. B}\ }\textbf {\bibinfo {volume} {86}},\ \bibinfo
  {pages} {125114} (\bibinfo {year} {2012})}\BibitemShut {NoStop}%
\bibitem [{\citenamefont {Bychkov}\ \emph {et~al.}(1966)\citenamefont
  {Bychkov}, \citenamefont {Gor'kov},\ and\ \citenamefont
  {Dzyaloshinski}}]{e_023_03_0489}%
  \BibitemOpen
  \bibfield  {author} {\bibinfo {author} {\bibfnamefont {Y.~A.}\ \bibnamefont
  {Bychkov}}, \bibinfo {author} {\bibfnamefont {L.~P.}\ \bibnamefont
  {Gor'kov}},\ and\ \bibinfo {author} {\bibfnamefont {I.~E.}\ \bibnamefont
  {Dzyaloshinski}},\ }\bibinfo {title} {Possibility of superconductivity type
  phenomena in a one-dimensional system},\ \href
  {http://jetp.ras.ru/cgi-bin/e/index/r/50/3/p738?a=list} {\bibfield  {journal}
  {\bibinfo  {journal} {Journal of Experimental and Theoretical Physics}\
  }\textbf {\bibinfo {volume} {50}},\ \bibinfo {pages} {738} (\bibinfo {year}
  {1966})}\BibitemShut {NoStop}%
\bibitem [{\citenamefont {Janis}(1998)}]{Janis_1998}%
  \BibitemOpen
  \bibfield  {author} {\bibinfo {author} {\bibfnamefont {V.}~\bibnamefont
  {Janis}},\ }\bibinfo {title} {The hubbard model at intermediate coupling:
  renormalization of the interaction strength},\ \href
  {https://doi.org/10.1088/0953-8984/10/13/010} {\bibfield  {journal} {\bibinfo
   {journal} {Journal of Physics: Condensed Matter}\ }\textbf {\bibinfo
  {volume} {10}},\ \bibinfo {pages} {2915} (\bibinfo {year}
  {1998})}\BibitemShut {NoStop}%
\bibitem [{\citenamefont {Jani\ifmmode~\check{s}\else
  \v{s}\fi{}}(1999)}]{PhysRevB.60.11345}%
  \BibitemOpen
  \bibfield  {author} {\bibinfo {author} {\bibfnamefont {V.}~\bibnamefont
  {Jani\ifmmode~\check{s}\else \v{s}\fi{}}},\ }\bibinfo {title} {Stability of
  self-consistent solutions for the hubbard model at intermediate and strong
  coupling},\ \href {https://doi.org/10.1103/PhysRevB.60.11345} {\bibfield
  {journal} {\bibinfo  {journal} {Phys. Rev. B}\ }\textbf {\bibinfo {volume}
  {60}},\ \bibinfo {pages} {11345} (\bibinfo {year} {1999})}\BibitemShut
  {NoStop}%
\bibitem [{\citenamefont {Kusunose}(2010)}]{doi:10.1143/JPSJ.79.094707}%
  \BibitemOpen
  \bibfield  {author} {\bibinfo {author} {\bibfnamefont {H.}~\bibnamefont
  {Kusunose}},\ }\bibinfo {title} {Self-consistent fluctuation theory for
  strongly correlated electron systems},\ \href
  {https://doi.org/10.1143/JPSJ.79.094707} {\bibfield  {journal} {\bibinfo
  {journal} {Journal of the Physical Society of Japan}\ }\textbf {\bibinfo
  {volume} {79}},\ \bibinfo {pages} {094707} (\bibinfo {year} {2010})},\
  \Eprint {https://arxiv.org/abs/https://doi.org/10.1143/JPSJ.79.094707}
  {https://doi.org/10.1143/JPSJ.79.094707} \BibitemShut {NoStop}%
\bibitem [{\citenamefont {Jani\ifmmode~\check{s}\else \v{s}\fi{}}\ and\
  \citenamefont {Augustinsk\'y}(2007)}]{PhysRevB.75.165108}%
  \BibitemOpen
  \bibfield  {author} {\bibinfo {author} {\bibfnamefont {V.}~\bibnamefont
  {Jani\ifmmode~\check{s}\else \v{s}\fi{}}}\ and\ \bibinfo {author}
  {\bibfnamefont {P.}~\bibnamefont {Augustinsk\'y}},\ }\bibinfo {title}
  {Analytic impurity solver with kondo strong-coupling asymptotics},\ \href
  {https://doi.org/10.1103/PhysRevB.75.165108} {\bibfield  {journal} {\bibinfo
  {journal} {Phys. Rev. B}\ }\textbf {\bibinfo {volume} {75}},\ \bibinfo
  {pages} {165108} (\bibinfo {year} {2007})}\BibitemShut {NoStop}%
\bibitem [{\citenamefont {Augustinsk\'y}\ and\ \citenamefont
  {Jani\ifmmode~\check{s}\else \v{s}\fi{}}(2011)}]{PhysRevB.83.035114}%
  \BibitemOpen
  \bibfield  {author} {\bibinfo {author} {\bibfnamefont {P.}~\bibnamefont
  {Augustinsk\'y}}\ and\ \bibinfo {author} {\bibfnamefont {V.}~\bibnamefont
  {Jani\ifmmode~\check{s}\else \v{s}\fi{}}},\ }\bibinfo {title} {Multiorbital
  simplified parquet equations for strongly correlated electrons},\ \href
  {https://doi.org/10.1103/PhysRevB.83.035114} {\bibfield  {journal} {\bibinfo
  {journal} {Phys. Rev. B}\ }\textbf {\bibinfo {volume} {83}},\ \bibinfo
  {pages} {035114} (\bibinfo {year} {2011})}\BibitemShut {NoStop}%
\bibitem [{\citenamefont {Mizuno}\ \emph {et~al.}(2021)\citenamefont {Mizuno},
  \citenamefont {Ochi},\ and\ \citenamefont {Kuroki}}]{PhysRevB.104.035160}%
  \BibitemOpen
  \bibfield  {author} {\bibinfo {author} {\bibfnamefont {R.}~\bibnamefont
  {Mizuno}}, \bibinfo {author} {\bibfnamefont {M.}~\bibnamefont {Ochi}},\ and\
  \bibinfo {author} {\bibfnamefont {K.}~\bibnamefont {Kuroki}},\ }\bibinfo
  {title} {Development of an efficient impurity solver in dynamical mean field
  theory for multiband systems: Iterative perturbation theory combined with
  parquet equations},\ \href {https://doi.org/10.1103/PhysRevB.104.035160}
  {\bibfield  {journal} {\bibinfo  {journal} {Phys. Rev. B}\ }\textbf {\bibinfo
  {volume} {104}},\ \bibinfo {pages} {035160} (\bibinfo {year}
  {2021})}\BibitemShut {NoStop}%
\bibitem [{Note1()}]{Note1}%
  \BibitemOpen
  \bibinfo {note} {Although the vertices represented by squares depend on the
  three frequencies, we can expect that their constant part mainly contributes
  to the global frequency structure of the full-vertex we focus on in this
  study. In other words, since they have frequency dependencies only near the
  origin of $(\omega , \omega ', \nu )$-space, their frequency-dependent parts
  have contributions only in the vicinity of the origin, and have little effect
  on the global frequency structure of the full-vertex.}\BibitemShut {Stop}%
\bibitem [{\citenamefont {Yosida}\ and\ \citenamefont
  {Yamada}(1970)}]{doi:10.1143/PTPS.46.244}%
  \BibitemOpen
  \bibfield  {author} {\bibinfo {author} {\bibfnamefont {K.}~\bibnamefont
  {Yosida}}\ and\ \bibinfo {author} {\bibfnamefont {K.}~\bibnamefont
  {Yamada}},\ }\bibinfo {title} {Perturbation expansion for the anderson
  hamiltonian},\ \href {https://doi.org/10.1143/PTPS.46.244} {\bibfield
  {journal} {\bibinfo  {journal} {Progress of Theoretical Physics Supplement}\
  }\textbf {\bibinfo {volume} {46}},\ \bibinfo {pages} {244} (\bibinfo {year}
  {1970})}\BibitemShut {NoStop}%
\bibitem [{\citenamefont
  {Yamada}(1975{\natexlab{a}})}]{doi:10.1143/PTP.53.970}%
  \BibitemOpen
  \bibfield  {author} {\bibinfo {author} {\bibfnamefont {K.}~\bibnamefont
  {Yamada}},\ }\bibinfo {title} {Perturbation expansion for the anderson
  hamiltonian. ii},\ \href {https://doi.org/10.1143/PTP.53.970} {\bibfield
  {journal} {\bibinfo  {journal} {Progress of Theoretical Physics}\ }\textbf
  {\bibinfo {volume} {53}},\ \bibinfo {pages} {970} (\bibinfo {year}
  {1975}{\natexlab{a}})}\BibitemShut {NoStop}%
\bibitem [{\citenamefont {Yosida}\ and\ \citenamefont
  {Yamada}(1975)}]{doi:10.1143/PTP.53.1286}%
  \BibitemOpen
  \bibfield  {author} {\bibinfo {author} {\bibfnamefont {K.}~\bibnamefont
  {Yosida}}\ and\ \bibinfo {author} {\bibfnamefont {K.}~\bibnamefont
  {Yamada}},\ }\bibinfo {title} {Perturbation expansion for the anderson
  hamiltonian. iii},\ \href {https://doi.org/10.1143/PTP.53.1286} {\bibfield
  {journal} {\bibinfo  {journal} {Progress of Theoretical Physics}\ }\textbf
  {\bibinfo {volume} {53}},\ \bibinfo {pages} {1286} (\bibinfo {year}
  {1975})}\BibitemShut {NoStop}%
\bibitem [{\citenamefont {Yamada}(1975{\natexlab{b}})}]{Yamada4}%
  \BibitemOpen
  \bibfield  {author} {\bibinfo {author} {\bibfnamefont {K.}~\bibnamefont
  {Yamada}},\ }\bibinfo {title} {Perturbation expansion for the anderson
  hamiltonian. iv},\ \href {https://doi.org/10.1143/PTP.54.316} {\bibfield
  {journal} {\bibinfo  {journal} {Progress of Theoretical Physics}\ }\textbf
  {\bibinfo {volume} {54}},\ \bibinfo {pages} {316} (\bibinfo {year}
  {1975}{\natexlab{b}})}\BibitemShut {NoStop}%
\bibitem [{\citenamefont {Georges}\ and\ \citenamefont
  {Kotliar}(1992)}]{PhysRevB.45.6479}%
  \BibitemOpen
  \bibfield  {author} {\bibinfo {author} {\bibfnamefont {A.}~\bibnamefont
  {Georges}}\ and\ \bibinfo {author} {\bibfnamefont {G.}~\bibnamefont
  {Kotliar}},\ }\bibinfo {title} {Hubbard model in infinite dimensions},\ \href
  {https://doi.org/10.1103/PhysRevB.45.6479} {\bibfield  {journal} {\bibinfo
  {journal} {Phys. Rev. B}\ }\textbf {\bibinfo {volume} {45}},\ \bibinfo
  {pages} {6479} (\bibinfo {year} {1992})}\BibitemShut {NoStop}%
\bibitem [{\citenamefont {Rubtsov}\ and\ \citenamefont
  {Lichtenstein}(2004)}]{Rubtsov2004}%
  \BibitemOpen
  \bibfield  {author} {\bibinfo {author} {\bibfnamefont {A.~N.}\ \bibnamefont
  {Rubtsov}}\ and\ \bibinfo {author} {\bibfnamefont {A.~I.}\ \bibnamefont
  {Lichtenstein}},\ }\bibinfo {title} {Continuous-time quantum monte carlo
  method for fermions: Beyond auxiliary field framework},\ \href
  {https://doi.org/10.1134/1.1800216} {\bibfield  {journal} {\bibinfo
  {journal} {Journal of Experimental and Theoretical Physics Letters}\ }\textbf
  {\bibinfo {volume} {80}},\ \bibinfo {pages} {61} (\bibinfo {year}
  {2004})}\BibitemShut {NoStop}%
\bibitem [{\citenamefont {Nukala}\ \emph {et~al.}(2009)\citenamefont {Nukala},
  \citenamefont {Maier}, \citenamefont {Summers}, \citenamefont {Alvarez},\
  and\ \citenamefont {Schulthess}}]{PhysRevB.80.195111}%
  \BibitemOpen
  \bibfield  {author} {\bibinfo {author} {\bibfnamefont {P.~K. V.~V.}\
  \bibnamefont {Nukala}}, \bibinfo {author} {\bibfnamefont {T.~A.}\
  \bibnamefont {Maier}}, \bibinfo {author} {\bibfnamefont {M.~S.}\ \bibnamefont
  {Summers}}, \bibinfo {author} {\bibfnamefont {G.}~\bibnamefont {Alvarez}},\
  and\ \bibinfo {author} {\bibfnamefont {T.~C.}\ \bibnamefont {Schulthess}},\
  }\bibinfo {title} {Fast update algorithm for the quantum monte carlo
  simulation of the hubbard model},\ \href
  {https://doi.org/10.1103/PhysRevB.80.195111} {\bibfield  {journal} {\bibinfo
  {journal} {Phys. Rev. B}\ }\textbf {\bibinfo {volume} {80}},\ \bibinfo
  {pages} {195111} (\bibinfo {year} {2009})}\BibitemShut {NoStop}%
\bibitem [{\citenamefont {Gull}\ \emph {et~al.}(2011)\citenamefont {Gull},
  \citenamefont {Staar}, \citenamefont {Fuchs}, \citenamefont {Nukala},
  \citenamefont {Summers}, \citenamefont {Pruschke}, \citenamefont
  {Schulthess},\ and\ \citenamefont {Maier}}]{PhysRevB.83.075122}%
  \BibitemOpen
  \bibfield  {author} {\bibinfo {author} {\bibfnamefont {E.}~\bibnamefont
  {Gull}}, \bibinfo {author} {\bibfnamefont {P.}~\bibnamefont {Staar}},
  \bibinfo {author} {\bibfnamefont {S.}~\bibnamefont {Fuchs}}, \bibinfo
  {author} {\bibfnamefont {P.}~\bibnamefont {Nukala}}, \bibinfo {author}
  {\bibfnamefont {M.~S.}\ \bibnamefont {Summers}}, \bibinfo {author}
  {\bibfnamefont {T.}~\bibnamefont {Pruschke}}, \bibinfo {author}
  {\bibfnamefont {T.~C.}\ \bibnamefont {Schulthess}},\ and\ \bibinfo {author}
  {\bibfnamefont {T.}~\bibnamefont {Maier}},\ }\bibinfo {title} {Submatrix
  updates for the continuous-time auxiliary-field algorithm},\ \href
  {https://doi.org/10.1103/PhysRevB.83.075122} {\bibfield  {journal} {\bibinfo
  {journal} {Phys. Rev. B}\ }\textbf {\bibinfo {volume} {83}},\ \bibinfo
  {pages} {075122} (\bibinfo {year} {2011})}\BibitemShut {NoStop}%
\bibitem [{\citenamefont {Nomura}\ \emph {et~al.}(2014)\citenamefont {Nomura},
  \citenamefont {Sakai},\ and\ \citenamefont {Arita}}]{PhysRevB.89.195146}%
  \BibitemOpen
  \bibfield  {author} {\bibinfo {author} {\bibfnamefont {Y.}~\bibnamefont
  {Nomura}}, \bibinfo {author} {\bibfnamefont {S.}~\bibnamefont {Sakai}},\ and\
  \bibinfo {author} {\bibfnamefont {R.}~\bibnamefont {Arita}},\ }\bibinfo
  {title} {Multiorbital cluster dynamical mean-field theory with an improved
  continuous-time quantum monte carlo algorithm},\ \href
  {https://doi.org/10.1103/PhysRevB.89.195146} {\bibfield  {journal} {\bibinfo
  {journal} {Phys. Rev. B}\ }\textbf {\bibinfo {volume} {89}},\ \bibinfo
  {pages} {195146} (\bibinfo {year} {2014})}\BibitemShut {NoStop}%
\bibitem [{\citenamefont {Seth}\ \emph {et~al.}(2016)\citenamefont {Seth},
  \citenamefont {Krivenko}, \citenamefont {Ferrero},\ and\ \citenamefont
  {Parcollet}}]{Seth2016274}%
  \BibitemOpen
  \bibfield  {author} {\bibinfo {author} {\bibfnamefont {P.}~\bibnamefont
  {Seth}}, \bibinfo {author} {\bibfnamefont {I.}~\bibnamefont {Krivenko}},
  \bibinfo {author} {\bibfnamefont {M.}~\bibnamefont {Ferrero}},\ and\ \bibinfo
  {author} {\bibfnamefont {O.}~\bibnamefont {Parcollet}},\ }\bibinfo {title}
  {Triqs/cthyb: A continuous-time quantum monte carlo hybridisation expansion
  solver for quantum impurity problems},\ \href
  {https://doi.org/https://doi.org/10.1016/j.cpc.2015.10.023} {\bibfield
  {journal} {\bibinfo  {journal} {Computer Physics Communications}\ }\textbf
  {\bibinfo {volume} {200}},\ \bibinfo {pages} {274 } (\bibinfo {year}
  {2016})}\BibitemShut {NoStop}%
\bibitem [{\citenamefont {Gull}(2008)}]{Gull_phdthesis}%
  \BibitemOpen
  \bibfield  {author} {\bibinfo {author} {\bibfnamefont {E.}~\bibnamefont
  {Gull}},\ }\href@noop {} {Ph.D. thesis},\ \bibinfo  {school} {ETH Z\"{u}rich}
  (\bibinfo {year} {2008})\BibitemShut {NoStop}%
\bibitem [{\citenamefont {Boehnke}(2015)}]{lewin_thesis}%
  \BibitemOpen
  \bibfield  {author} {\bibinfo {author} {\bibfnamefont {L.~V.}\ \bibnamefont
  {Boehnke}},\ }\emph {\bibinfo {title} {{Susceptibilities in materials with
  multiple strongly correlated orbitals}}},\ \href
  {http://ediss.sub.uni-hamburg.de/volltexte/2015/7325/pdf/Dissertation.pdf}
  {Ph.D. thesis},\ \bibinfo  {school} {Universit\"{a}t Hamburg} (\bibinfo
  {year} {2015})\BibitemShut {NoStop}%
\bibitem [{\citenamefont {Boehnke}\ \emph {et~al.}(2011)\citenamefont
  {Boehnke}, \citenamefont {Hafermann}, \citenamefont {Ferrero}, \citenamefont
  {Lechermann},\ and\ \citenamefont {Parcollet}}]{triqs_ctqmc_solver_legendre}%
  \BibitemOpen
  \bibfield  {author} {\bibinfo {author} {\bibfnamefont {L.}~\bibnamefont
  {Boehnke}}, \bibinfo {author} {\bibfnamefont {H.}~\bibnamefont {Hafermann}},
  \bibinfo {author} {\bibfnamefont {M.}~\bibnamefont {Ferrero}}, \bibinfo
  {author} {\bibfnamefont {F.}~\bibnamefont {Lechermann}},\ and\ \bibinfo
  {author} {\bibfnamefont {O.}~\bibnamefont {Parcollet}},\ }\bibinfo {title}
  {Orthogonal polynomial representation of imaginary-time green's functions},\
  \href {https://doi.org/10.1103/PhysRevB.84.075145} {\bibfield  {journal}
  {\bibinfo  {journal} {Phys. Rev. B}\ }\textbf {\bibinfo {volume} {84}},\
  \bibinfo {pages} {075145} (\bibinfo {year} {2011})}\BibitemShut {NoStop}%
\bibitem [{\citenamefont {L\"auchli}\ and\ \citenamefont
  {Werner}(2009)}]{PhysRevB.80.235117}%
  \BibitemOpen
  \bibfield  {author} {\bibinfo {author} {\bibfnamefont {A.~M.}\ \bibnamefont
  {L\"auchli}}\ and\ \bibinfo {author} {\bibfnamefont {P.}~\bibnamefont
  {Werner}},\ }\bibinfo {title} {Krylov implementation of the hybridization
  expansion impurity solver and application to 5-orbital models},\ \href
  {https://doi.org/10.1103/PhysRevB.80.235117} {\bibfield  {journal} {\bibinfo
  {journal} {Phys. Rev. B}\ }\textbf {\bibinfo {volume} {80}},\ \bibinfo
  {pages} {235117} (\bibinfo {year} {2009})}\BibitemShut {NoStop}%
\bibitem [{\citenamefont {S\'emon}\ \emph {et~al.}(2014)\citenamefont
  {S\'emon}, \citenamefont {Yee}, \citenamefont {Haule},\ and\ \citenamefont
  {Tremblay}}]{PhysRevB.90.075149}%
  \BibitemOpen
  \bibfield  {author} {\bibinfo {author} {\bibfnamefont {P.}~\bibnamefont
  {S\'emon}}, \bibinfo {author} {\bibfnamefont {C.-H.}\ \bibnamefont {Yee}},
  \bibinfo {author} {\bibfnamefont {K.}~\bibnamefont {Haule}},\ and\ \bibinfo
  {author} {\bibfnamefont {A.-M.~S.}\ \bibnamefont {Tremblay}},\ }\bibinfo
  {title} {Lazy skip-lists: An algorithm for fast hybridization-expansion
  quantum monte carlo},\ \href {https://doi.org/10.1103/PhysRevB.90.075149}
  {\bibfield  {journal} {\bibinfo  {journal} {Phys. Rev. B}\ }\textbf {\bibinfo
  {volume} {90}},\ \bibinfo {pages} {075149} (\bibinfo {year}
  {2014})}\BibitemShut {NoStop}%
\bibitem [{\citenamefont {Shinaoka}\ \emph {et~al.}(2017)\citenamefont
  {Shinaoka}, \citenamefont {Assaad}, \citenamefont {Bl{\"u}mer},\ and\
  \citenamefont {Werner}}]{Shinaoka2017}%
  \BibitemOpen
  \bibfield  {author} {\bibinfo {author} {\bibfnamefont {H.}~\bibnamefont
  {Shinaoka}}, \bibinfo {author} {\bibfnamefont {F.}~\bibnamefont {Assaad}},
  \bibinfo {author} {\bibfnamefont {N.}~\bibnamefont {Bl{\"u}mer}},\ and\
  \bibinfo {author} {\bibfnamefont {P.}~\bibnamefont {Werner}},\ }\bibinfo
  {title} {Quantum monte carlo impurity solvers for multi-orbital problems and
  frequency-dependent interactions},\ \href
  {https://doi.org/10.1140/epjst/e2017-70050-x} {\bibfield  {journal} {\bibinfo
   {journal} {The European Physical Journal Special Topics}\ }\textbf {\bibinfo
  {volume} {226}},\ \bibinfo {pages} {2499} (\bibinfo {year}
  {2017})}\BibitemShut {NoStop}%
\bibitem [{\citenamefont {Otsuki}\ \emph {et~al.}(2019)\citenamefont {Otsuki},
  \citenamefont {Yoshimi}, \citenamefont {Shinaoka},\ and\ \citenamefont
  {Nomura}}]{PhysRevB.99.165134}%
  \BibitemOpen
  \bibfield  {author} {\bibinfo {author} {\bibfnamefont {J.}~\bibnamefont
  {Otsuki}}, \bibinfo {author} {\bibfnamefont {K.}~\bibnamefont {Yoshimi}},
  \bibinfo {author} {\bibfnamefont {H.}~\bibnamefont {Shinaoka}},\ and\
  \bibinfo {author} {\bibfnamefont {Y.}~\bibnamefont {Nomura}},\ }\bibinfo
  {title} {Strong-coupling formula for momentum-dependent susceptibilities in
  dynamical mean-field theory},\ \href
  {https://doi.org/10.1103/PhysRevB.99.165134} {\bibfield  {journal} {\bibinfo
  {journal} {Phys. Rev. B}\ }\textbf {\bibinfo {volume} {99}},\ \bibinfo
  {pages} {165134} (\bibinfo {year} {2019})}\BibitemShut {NoStop}%
\bibitem [{\citenamefont {Stepanov}\ \emph {et~al.}(2019)\citenamefont
  {Stepanov}, \citenamefont {Harkov},\ and\ \citenamefont
  {Lichtenstein}}]{PhysRevB.100.205115}%
  \BibitemOpen
  \bibfield  {author} {\bibinfo {author} {\bibfnamefont {E.~A.}\ \bibnamefont
  {Stepanov}}, \bibinfo {author} {\bibfnamefont {V.}~\bibnamefont {Harkov}},\
  and\ \bibinfo {author} {\bibfnamefont {A.~I.}\ \bibnamefont {Lichtenstein}},\
  }\bibinfo {title} {Consistent partial bosonization of the extended hubbard
  model},\ \href {https://doi.org/10.1103/PhysRevB.100.205115} {\bibfield
  {journal} {\bibinfo  {journal} {Phys. Rev. B}\ }\textbf {\bibinfo {volume}
  {100}},\ \bibinfo {pages} {205115} (\bibinfo {year} {2019})}\BibitemShut
  {NoStop}%
\end{thebibliography}%

\end{document}